\documentclass[useAMS,usenatbib]{mn2e}

\usepackage{amssymb}
\usepackage{times}
\usepackage{graphicx} 

\title[Foreground simulations for the LOFAR - EoR
Experiment]{Foreground simulations for the LOFAR - Epoch of
Reionization Experiment}

\author[Jeli{\'c} et al. ]{ V. Jeli{\'c}$^{1}$\thanks{E-mail:
vjelic@astro.rug.nl}, S. Zaroubi$^{1}$, P. Labropoulos$^{1}$,
R.M. Thomas$^{1}$, G. Bernardi$^{1}$, \newauthor M.A. Brentjens$^{2}$,
A.G. de Bruyn$^{1,2}$, B. Ciardi$^{3}$, G. Harker$^{1}$, L.V.E.
Koopmans,$^{1}$,\newauthor V.N. Pandey$^{1}$, J.  Schaye$^{4}$,
S. Yatawatta$^{1}$\\ $^{1}$Kapteyn Astronomical Institute, University
of Groningen, P.O. Box 800, 9700 AV Groningen, the Netherlands\\
$^{2}$ASTRON, Postbus 2, 7990 AA Dwingeloo, the Netherlands\\
$^{3}$Max-Planck Institute for Astrophysics,
Karl-Schwarzschild-Stra\ss e 1, 85748 Garching, Germany\\ $^{4}$Leiden
Observatory, Leiden University, PO Box 9513, 2300 RA Leiden, the
Netherlands}

\begin{document}


\pagerange{\pageref{firstpage}--\pageref{lastpage}} \pubyear{2008}

\maketitle 

\label{firstpage}

\begin{abstract}
Future high redshift 21-cm experiments will suffer from a high degree
of contamination, due both to astrophysical foregrounds and to
non-astrophysical and instrumental effects.  In order to reliably
extract the cosmological signal from the observed data, it is
essential to understand very well all data components and their
influence on the extracted signal. Here we present simulated
astrophysical foregrounds datacubes and discuss their possible
statistical effects on the data.  The foreground maps are produced
assuming $5^\circ\times 5^\circ$ windows that match those expected to
be observed by the LOFAR Epoch-of-Reionization (EoR) key science
project. We show that with the expected LOFAR-EoR sky and receiver
noise levels, which amount to $\approx 52\mathrm{mK}$ at 150~{\rm MHz}
after 400 hours of total observing time, a simple polynomial fit
allows a statistical reconstruction of the signal.  We also show that
the polynomial fitting will work for maps with realistic yet idealised
instrument response, i.e., a response that includes only a uniform uv
coverage as a function of frequency and ignores many other
uncertainties.  Polarized galactic synchrotron maps that include
internal polarization and a number of Faraday screens along the line
of sight are also simulated. The importance of these stems from the
fact that the LOFAR instrument, in common with all current
interferometric EoR experiments has an instrumentally polarized
response.

\end{abstract}


\section{Introduction}
\label{sec:introduction}

The Epoch of Reionization (hereafter, EoR), which marks the end of the
Universe's `Dark Ages', is one of the least explored epochs in cosmic
evolution.  Currently, there are two main observational constraints on
the EoR. The first is the sudden jump in the Lyman-$\alpha$ optical
depth in the Gunn-Peterson troughs \citep{gunn65} observed in the
Sloan Digital Sky Survey quasar spectra \citep{becker01, fan01,
pentericci02, white03, fan06}, marking a lower limit to the redshift
at which the Universe became completely ionized.  The second
constraint comes from the fifth year WMAP satellite data on the
temperature and polarization anisotropies of the cosmic microwave
background (CMB) \citep{spergel07, page07} which gives an integral
constraint on the Thomson optical depth experienced by the CMB photons
since the EoR. However, both of these observational methods provide
limited information on the reionization process.

The redshifted 21-cm hyperfine transition line of neutral hydrogen is
the most promising and immediately accessible method for probing the
intergalactic medium (IGM) during reionization
\citep[e.g.~][]{field58, field59, scott90, kumar95, madau97}.  Recent
years have witnessed a flurry of theoretical activities to predict
reionization sources and their impact on the IGM
\citep[e.g.~][]{barkana01, loeb01, ciardi03a, ciardi03b, bromm04,
iliev07, zaroubi07, thomas07}. Measurements of the 21-cm
signal can also help to constrain the cosmological parameters
independently \citep{mcquinn06}.

Future telescopes like LOFAR\footnote{http://www.lofar.org},
MWA\footnote{http://www.haystack.mit.edu/ast/arrays/mwa},
21CMA\footnote{http://web.phys.cmu.edu/~past/} and
SKA\footnote{http://www.skatelescope.org} are being designed to study
the redshifted 21-cm signal from the EoR. A successful detection of
this signal will help us derive the nature of the first sources and
their impact on the surrounding IGM.  

Unfortunately however, the cosmological EoR signal is contaminated by
a slew of astrophysical and non-astrophysical components. Typically,
the contamination level is orders of magnitude larger than the
cosmological 21-cm signal. Thus, the primary challenge of the EoR
observations will be the accurate modelling of the various data
components -- foregrounds, instrumental response, ionospheric
disturbances, to name a few -- which is essential to develop a robust
signal extraction scheme.

For the foregrounds, there are currently no available data in the
115-180~{\rm MHz} frequency range and 4 arcmin resolution at high
Galactic latitude that would allow accurate modelling of the LOFAR-EoR
foregrounds. Therefore, one has to rely on the available relevant data
and extrapolate, based on theoretical arguments, into the frequency
range and resolution observed by LOFAR. However, recently
\citet{ali08} used $153~{\rm MHz}$ obervations with Giant Meter Wave
Radio Telescope to characterize the statistical properties --
visibility correlation function -- of the foregrounds. This paper
focuses on simulating the galactic and extragalactic foregrounds that
dominate the sky at frequencies of interest for the LOFAR-EoR
experiment ($115$--$215~{\rm MHz}$). The main foreground components
are: Galactic synchrotron emission from diffuse and localised sources,
Galactic thermal (free-free) emission and integrated emission from
extragalactic sources (like radio galaxies and clusters). The dominant
component of the foregrounds is the Galactic synchrotron emission
($\sim$70 per cent). The extragalactic emission contributes $\sim$27
per cent and Galactic free-free emission $\sim$1 per cent
\citep{shaver99}. Although the difference between the mean amplitude
of the EoR signal and the foregrounds is expected to be 4-5 orders of
magnitude, an interferometer like LOFAR measures only the fluctuations
which in this case are expected to be different by `only' three orders
of magnitude.

Various authors have studied the foregrounds in the context the EoR
measurements. \citet{shaver99} have studied the diffuse synchrotron
and free-free emission from our Galaxy and extragalactic sources;
\citet{dimatteo02} and \citet{dimatteo04} have considered emission
from unresolved extragalactic sources at low radio frequencies; and
\citet{oh03} and \citet{cooray04} studied the effect of free-free
emission from extragalactic haloes. Over the years, several methods
have been explored to filter out the foregrounds. Most of the methods
rely on the relative smoothness in the frequency of the foregrounds,
with respect to the signal \citep{shaver99, dimatteo02,
zaldarriaga04, morales06, wang06, gleser07}.

\citet{santos05} have studied the foregrounds for the EoR experiment
and their influence on the measurement of the 21-cm signal. In their
multi-frequency analysis of the power spectra, they considered four
types of foregrounds: Galactic diffuse synchrotron emission;
Galactic free-free emission; extragalactic free-free emission; and
extragalactic point sources. They showed that foregrounds cleaning is
aided by the large scale angular correlation, especially of the
extragalactic point sources, which facilitates signal extraction to a
level suitable for the EoR experiments.

The current study is part of the general effort undertaken by the
LOFAR-EoR key science project to produce simulated data cubes.  The
pipeline under construction will simulate the LOFAR-EoR data cube that
includes the simulated cosmological 21-cm signal, the galactic and
extragalactic foregrounds, ionospheric effects, radio frequency
interferences (RFIs) and the instrumental response.  These datacubes
will be used to design the observational strategy and test our
signal-processing methods. Our main concern in this paper is the
simulation of the galactic and extragalactic foregrounds.

Recently, a study by \citet{gleser07} has been conducted along lines
similar to parts of the current paper. The authors test a certain
signal extraction algorithm on simulated foregrounds maps in which
they take most of the relevant foregrounds into account. However,
there are many important differences between the two papers. First, in
the \citet{gleser07} study the assumption for the noise level in the
LOFAR-EoR project, as well as the other experiments, is at least an
order of magnitude too low. They assume 1 and 5~{\rm mK} noise models
whereas in reality the noise for the LOFAR-EoR experiment is about
50~{\rm mK}. They also present a simplified model of the Galactic
foregrounds that does not take into account all the spatial and
frequency correlations of the Galactic diffuse synchrotron emission
and underestimates that of the Galactic free-free emission, both of
which are very important. In contrast to them, we also present
polarized maps and introduce the LOFAR instrumental response and noise
in a realistic manner.

In the foregrounds simulations presented in this paper we choose a
different approach from previous groups, since our main aim is to
produce the simulations that will be part of the LOFAR-EoR data
pipeline. In this context our main aim is to produce foregrounds maps
in the angular and frequency range of the LOFAR-EoR experiment, i.e.
3D datacubes, and then use those simulations for testing the accuracy
of removal of the foregrounds. Section~\ref{sec:polarization} outlines
the importance of the polarized character of the foregrounds and how
to model the Stokes I, Q, and U polarization maps of the Galactic
synchrotron emission. Section~\ref{sec:inst} presents simulated
instrumental effects of the LOFAR telescope and their influence on the
foregrounds maps, and Section~\ref{sec:extraction} discusses a method
to extract the EoR signal from the foregrounds. The paper concludes
with a discussion and outlook (Section~\ref{sec:discussion and
outlook}).

\section{The Cosmological 21-cm signal}
\label{sec:signal}

In radio astronomy, where the Rayleigh-Jeans law is applicable, the
radiation intensity, $I(\nu)$ is expressed in terms of the brightness
temperature $T_{b}$, such that:
\begin{equation}
I(\nu) = \frac{2 \nu^2}{c^2} k T_b,
\end{equation}
where $\nu$ is the frequency, $c$ is the speed of light and $k$ is
Boltzmann's constant. The predicted differential brightness
temperature deviation of the cosmological 21-cm signal from the cosmic
microwave background radiation is given by \citep{field58, field59,
ciardi03}:
\begin{eqnarray}
\delta T_{\rm b} & = &26~\mathrm{mK}~x_{\rm HI}(1+\delta)
 \left(1-\frac{T_{\rm CMB}}{T_{\rm s}}\right) \left(\frac{\Omega_{\rm b}
 h^2}{0.02}\right) \nonumber\\ & &\left[\left(\frac{1 +
 z}{10}\right)\left(\frac{0.3}{\Omega_{\rm m}}\right)\right]^{1/2}.
\label{eq:tbright}
\end{eqnarray}
Here $T_{\rm s}$ is the spin temperature, $x_{\rm HI}$ is the neutral
hydrogen fraction, $\delta$ is the matter density contrast,
$\Omega_{\rm m}$ and $\Omega_{\rm b}$ are the mass and baryon density
in units of the critical density and $h=H_{0}/100$ \footnote{We assume
a $\Lambda$CDM Universe with $\Omega_{\rm b}=0.04$, $\Omega_{\rm
m}=0.26$, $\Omega_{\Lambda}=0.738$ and $H_{0}=70.8~{\rm k~ms^{-1}
Mpc^{-1}}$}.

In his seminal papers, \citet{field58, field59} used the quasi-static
approximation to calculate the spin temperature, $T_{s},$ as a
weighted average of the CMB, kinetic and colour temperature
\citep{wouthuysen52, field58}:
\begin{equation}
T_{s} = \frac{T_{CMB} +y_{kin} T_{kin} + y_\alpha
T_{\alpha}}{1+y_{kin}+y_\alpha},
\label{eq:tspin}
\end{equation}  
where $T_{CMB}$ is the CMB temperature and $y_{kin}$ and $y_\alpha$
are the kinetic and Lyman-$\alpha$ coupling terms, respectively. We
have assumed that the color temperature, $T_{\alpha}$, is equal to
$T_{kin}$. The kinetic coupling term increases with the kinetic
temperature, whereas the $y_\alpha$ coupling term is due to the
Lyman-$\alpha$ pumping, known also as the Wouthuysen-Field effect
\citep{wouthuysen52, field58}. The two coupling terms are dominant
under different conditions and in principle could be used to
distinguish between ionization sources, e.g., between first stars, for
which Lyman-$\alpha$ pumping is dominant, vs. first mini-quasars for
which X-ray photons and therefore heating is dominant \citep[see
e.g.,~][]{nusser05, kuhlen06, zaroubi07, thomas07}.

\begin{figure}
\centering
\vspace{-0.0cm}
\hspace{-0.5cm}
\includegraphics[width=.5\textwidth]{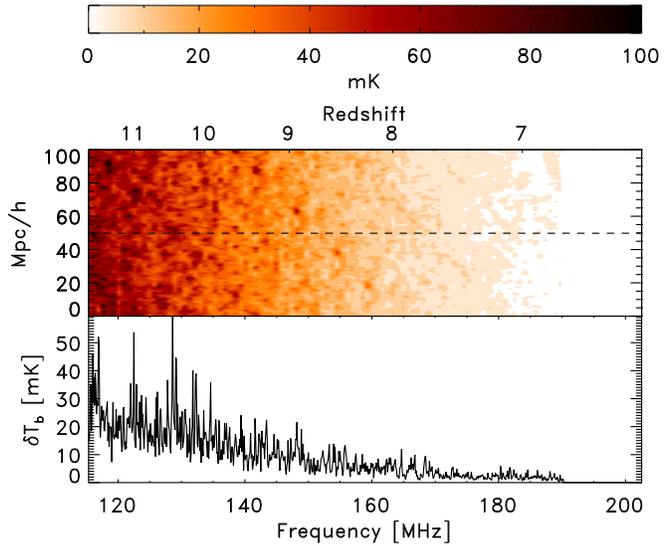}
\vspace{-0.5cm}
\caption{\emph{Simulated EoR signal assuming an exponential
form for the reionization history as decribed in the text and $T_{s}
\gg T_{CMB}$. The simulation box is $100~{\rm Mpc~h^{-1}}$ (comoving)
a side.  The upper panel shows the differential brightness temperature
in a slice along the redshift/frequency direction and another spatial
direction.  The lower panel shows the brightness temperature as a
function of redshift/frequency along a certain sight line (the dashed
line in the upper panel). The resolution along the frequency direction
is $10~{\rm kHz}$.}}
\label{fig:signal}
\end{figure}

The brightness temperature of the cosmological signal used in this
study is produced from a dark-matter-only N-body simulation. This
simulation is used to produce a cube of the cosmological signal, i.e.,
the density as a function of right ascesion, declination, and redshift
\citep[for more details see][in preparation]{thomas08}. Although
$T_{s}$ is calculated according to Eq.~\ref{eq:tspin}, we assume that
$T_s\gg T_{CMB}$. The reason for this assumption is that towards the
redshifts of interest for the experiment ($z=$6--12), the abundance of
Ly$\alpha$ photons in the Universe is sufficient to couple $T_s$ to
$T_{k}$ which is obviously much greater than $T_{CMB}$
\citep{ciardi03}. Hence from Eq.~\ref{eq:tbright}, $T_b$ follows the
cosmological density and $x_{HI}$. We further assume that along each
sight-line the neutral fraction follows the function $1/\left( 1 +
\exp (z-z_{reion})\right)$, where $z_{reion}$ for each pixel (or line
of sight) is set to $8.5\pm \delta_{z=10}$ and where $\delta_{z=10}$
is the density contrast at redshift 10. We used this approach to
randomize the reionization histories along different lines of sight
while preserving the spatial correlations of the cosmological
signals. In principle, this randomization could be drawn out of a
Gaussian distribution function. Redshift 10 here is an arbitrary
choice.  $z_{reion}$ along each line of sight varies in accordance
with the cosmological density along that line-of-sight at z=10 and has
a variance of unity centred at 8.5. Fig.~\ref{fig:signal} shows the
signal data cube that we use in order to test our foregrounds
filtering procedure.

Currently, a number of experiments (e.g., LOFAR, 21CMA, MWA and SKA)
are being designed to directly measure $\delta T_b$ of the HI 21-cm
hyperfine line and probe the physics of the reionization process by
observing the neutral fraction of the IGM as a function of
redshift. In this study, we focus on predictions for LOFAR, but our
conclusions could be easily applied to the other telescopes.

The LOFAR-EoR\footnote{For more information, see the LOFAR web site:
www.lofar.org and the LOFAR-EoR web site: www.astro.rug.nl/~LofarEoR.}
key project plans to measure the brightness fluctuations in the
frequency range of 115--190~{\rm MHz}, corresponding to redshift range
6-11.5 with spectral resolution of $\approx 1~\mathrm{M Hz}$ and
angular resolution of about $\approx 4$~{\rm arcmin}. A more detailed
discription of the LOFAR array will be given later in the paper when
the instrumental effects are discussed (Section~\ref{sec:inst}).

\section{Galactic foregrounds}
\label{sec:galactic}
The Galactic foregrounds have three main contributions. The first and
largest component is the Galactic diffuse synchrotron emission (GDSE),
which is the dominant foreground component in the frequency range of
the LOFAR-EoR experiment. The second component is radio synchrotron
emission from discrete sources, mostly supernova remnants (SNRs). The
third and last component is the free-free radio emission from
diffuse ionized gas. This component is the weakest of the three, yet it
still dominates over the cosmological component. Moreover, it has a
different spectral dependence, making it very imporant in testing the
signal extraction schemes that we have. In this section we describe
how we simulate the contribution of each of these components to the
total intensity. The polarized intensity simulations are described
later on.

\subsection{Galactic diffuse synchrotron emission (GDSE)}\label{gdse}
The GDSE originates from the interaction between the free electrons in
the interstellar medium and the Galactic magnetic field. Therefore
the observed GDSE intensity as a function of frequency, $I(\nu)$,
depends on the number density of emitting electrons, $N_{e}$, and the
Galactic magnetic field component perpendicular to the line of sight,
$B_{\perp}$:
\begin{equation}\label{synch}
  I(\nu)\sim N_{e}B^{(\gamma+1)/2}_{\perp}\nu^{-(\gamma-1)/2}
\end{equation}
where $\gamma$ is the electron spectral energy distribution power law
index \citep{pacholczyk70}.  The intensity of the synchrotron emission
as expressed in terms of the brightness temperature varies with
position and frequency and its spectrum is close to a featureless
power law $T_{b}\sim \nu^{\beta}$, where $\beta$ is the brightness
temperature spectral index, related to $\gamma$ by
$\beta=\textbf{-}(2+(\gamma-1)/2)$.

Observational data that are relevant to the LOFAR-EoR project are
scarce. \citet{landecker70} have produced an all sky map of the total
intensity of the GSDE at low radio frequencies at 150~{\rm MHz} with
5$^\circ$ resolution. The other Galactic survey relevant to the
LOFAR-EoR experiment is the 408~{\rm MHz} survey of \citet{haslam82}
with a resolution of $0.85^\circ$ and of \citet{reich88} at $1420~{\rm
MHz}$ with $0.95^\circ$ resolution. In the \citet{reich88} paper the
authors also assume a smooth power law change in the intensity as a
function of frequency which they calculate from their 1420~{\rm MHz}
and 408~{\rm MHz} maps.

At high Galactic latitudes the minimum brightness temperature of the
GDSE is about 20~K at 325~{\rm MHz} with variations of the order of 2
per cent on scales from 5--30~arcmin across the sky
\citep[][]{debruyn98}. At the same Galactic latitudes, the temperature
spectral index $\beta$ of the GDSE is about $-2.55$ at 100~{\rm MHz}
and steepens towards higher frequencies \citep[e.g.][]{reich88,
platania98}. Furthermore, the spectral index gradually changes with
position on the sky. This change appears to be caused by a variation
in the spectral index along the line of sight. An appropriate standard
deviation in the power law index, $\sigma_{\beta}$, in the frequency
range 100--200~{\rm MHz} appears to be of the order of $ \sim 0.1$
\citep{shaver99}. Recent data, collected around a galaxy cluster Abell
2255 using the WSRT telescope at 350~{\rm MHz}, indicate that the rms
of the brightness temperature at 3~arcmin resolution could be as low
as 0.1--0.3~{\rm K} (Pizzo and de Bruyn, private
communication). If extrapolated to 150~${\rm MHz}$ this result implies
that the rms in that region could be 1--2~${\rm K}$, which is an order
of magnitude smaller than the low resolution data suggest.

For the purpose of this paper we assume that the GDSE as a function of
frequency is well approximated by a power law within the limited
frequency range of 115--180~{\rm MHz}. This is a central assumption in
our simulation which is consistent with the general trend shown by the
available data, namely that the change in the frequency power law
index is gradual.  The values we choose for the power law index are
based on the high Galactic latitude regions in the \citet{haslam82}
and \citet{reich88} maps. The second assumption we make is that both
the intensity and power law index of the GDSE can be spatially
modelled as Gaussian random fields (GRFs). For the power spectrum of
GRFs we assume a power law with 2D index $\alpha=-2.7$. The standard
deviation of the GRFs is normalized to 0.4, assuming an angular scale
corresponding roughly to the field of view ($5^\circ$). This is
consistent with the value adopted by \citet{tegmark00},
\citet{giardino02} and \citet{santos05} for the angular power spectrum
index $\alpha$, where $C_{l} \sim l^{\alpha}$, $\alpha$ varies from
$-2.4$ to $-3$, and $l$ is the harmonic number.

In contrast to the previous authors \citep{tegmark00, giardino02,
santos05} who directly used the angular and frequency power spectrum
of the GDSE for their analysis, we simulate GDSE in four dimensions
(three spatial and one frequency), produce maps at each frequency and
then do our analysis on them. The four dimensional realisation
approach has the added benefit of enabling us to account for the
amplitude and temperature spectral index variations of the GDSE along
the line of sight ($z$-coordinate). We obtain the final map of the
GDSE at each frequency, $\nu$, by integrating the GDSE amplitude
($A(x,y,z,\nu)$) along the z-coordinate:
\begin{equation}\label{T_gdse}
  T_b(x,y,\nu)=C\int A(x,y,z,\nu)\mathrm{d}z
\end{equation}
where $T_b(x,y,\nu)$ is the brightness temperature of the GDSE as a
function of position and frequency and $C$ is a normalization
constant.  $A(x,y,z,\nu)$ is dimensionless and at each frequency is
defined by power law:
\begin{equation}\label{A_gdse}
  A(x,y,z,\nu)=A(x,y,z,\nu_{0})\left(
  \frac{\nu}{\nu_{0}}\right)^{\beta(x,y,z,\nu)}
\end{equation}
where $\nu_{0}$ is the reference frequency at which the normalisation
is done and $\beta(x,y,z,\nu)$ is the temperature spectral index as a
function of 3D position and frequency $\nu$. The power law index
$\beta$ has a weak frequency dependence, also as a power law.

$A(x,y,z,\nu_{0})$ and $\beta(x,y,z,\nu_{0})$ of the GDSE at the
normalization frequency $\nu_{0}$ are modelled spatially as two
Gaussian random fields with 3D power law spectrum $P(k)\sim
k^\delta$. Note that the absolute value of the 3D power law index $\delta$ is
$|\delta|=|\alpha|+1$ where $\alpha$ is the 2D power law index
mentioned above. $A(x,y,z,\nu_{0})$ and $\beta(x,y,z,\nu_{0})$ are
normalized according to observations (the Galactic surveys mentioned
above).

\begin{figure}
\centering
\vspace{-0.4cm}
\hspace{-0.5cm}
\includegraphics[width=.5\textwidth]{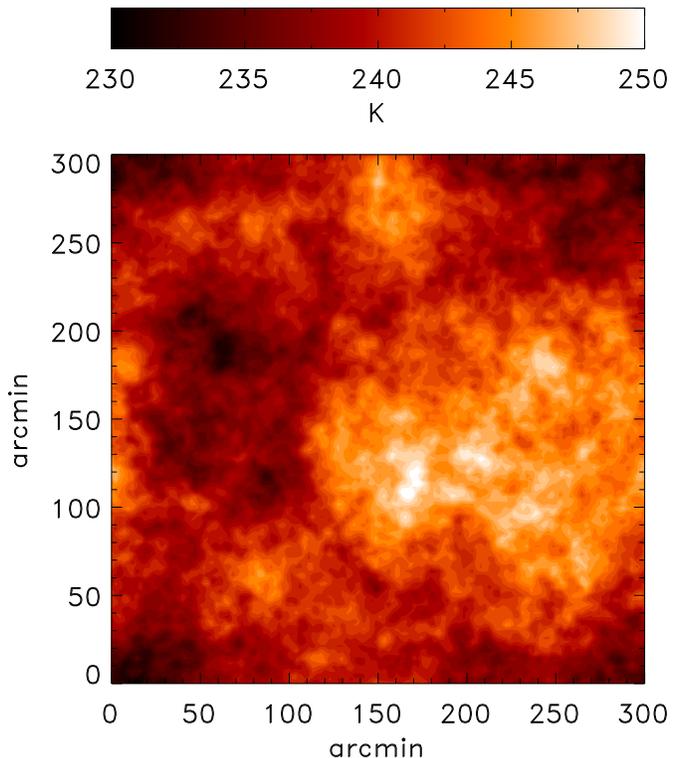}
\vspace{-1.0cm}
\caption{\emph{Simulated map of total intensity emission of Galactic
diffuse synchrotron emission with angular spectral index $\alpha=-2.7$
and frequency spectral index $\beta=-2.55$. The angular size of the map is
$5^{\circ}\times5^{\circ}$, with $\sim0.6'$ resolution. The colour bar
represents the brightness temperature $T_{b}$ of the Galactic diffuse
synchrotron emission in Kelvin at $120~{\rm MHz}$.}}
\label{fig:gse}
\end{figure}

For clarity, the steps we followed to produce the GDSE maps are
listed below:

\begin{enumerate}

\item Generate the same 3D Gaussian random field for both $A$ and
  $\beta$. The assumption here is that both fields have a correlated
  spatial distribution, which is supported by visual inspection of the
  high Galactic latitude portions of the~\citet{reich88} maps. We have
  also explored the possibility that $A$ and $\beta$ are independent;
  this has led to results very similar to the correlated case, and
  therefore we show only maps in which $A$ and $\beta$ are
  correlated.

\item Normalize the mean and standard deviation of $A(x,y,z,\nu_{0})$
  by integrating along the $z$ direction and setting the mean and
  standard deviation of $T_b(x,y,\nu_0)$ to match the observations
  (the Galactic surveys mentioned above). In other words we set the
  integration constant $C$ in Eq.~\ref{T_gdse}, in a way that the
  properties of the field $A(x,y,z,\nu_{0})$ after integration match
  the observed properties of $T_b(x,y,\nu_0)$.

\item Normalize the mean and standard deviation of $\beta (x,y,z,
  \nu_{0})$ according to observations.

\item Use Eq.~\ref{A_gdse} to calculate $A$ at each frequency.

\item Integrate along the $z$-coordinate to get the two-dimensional maps
of the GDSE brightness temperature at each frequency $\nu$
(Eq.~\ref{T_gdse}).
\end{enumerate}

For the purpose of this paper we simulate the GDSE on $512^3$ grid,
  where the x,y plane corresponds to angular size of
  $5^{\circ}\times5^{\circ}$ and z direction scales between 0--1 in
  dimensionless units. The amplitude, $A$, of the GDSE is normalized
  in the way described above to match $T_{b}(325{~\rm MHz})=20~{\rm K}
  \pm 2 \% $ \citep{debruyn98}, while $\beta$ is normalized at
  $100~{\rm MHz}$: $\beta=-2.55\pm 0.1$ \citep{shaver99}.

Fig.~\ref{fig:gse} shows a simulated map of the Galactic diffuse
synchrotron emission according to the procedure described above, at a
frequency of $120~$MHz with an angular size of
$5^{\circ}\times5^{\circ}$ on a $512^2$ grid. The mean brightness
temperature of the map is $T_{b}=~{253\rm K}$ with $\sigma={1.3\rm
K}$.

\begin{figure}
\centering
\hspace{-0.5cm}
\includegraphics[width=.5\textwidth]{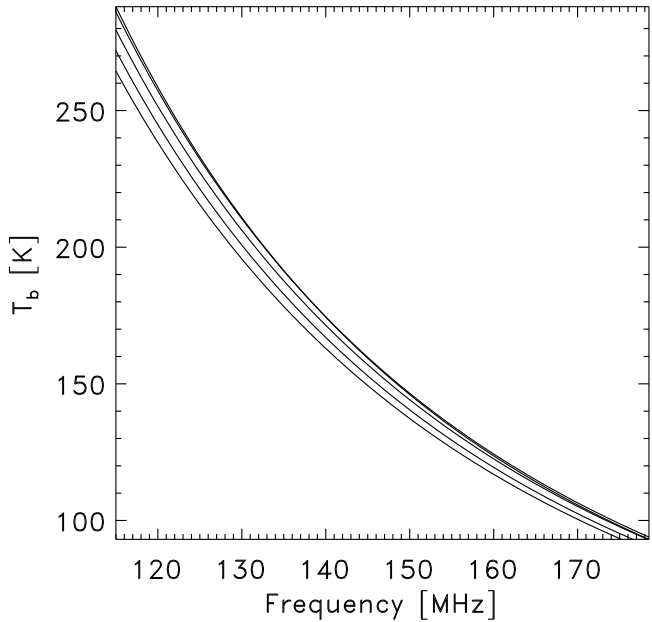}
\vspace{-0.5cm}
\caption{\emph{Brightness temperature of Galactic diffuse synchrotron
emission as a function of frequency, for five different lines of
sight. Each line of sight has a slightly different power law index
along the frequency direction as a result of the spatial and frequency
variations in the temperature spectral index.}}
\label{fig:gsefreq}
\end{figure}

In contrast to Fig.~\ref{fig:gse}, which shows the angular variations of
the GDSE at one frequency, Fig.~\ref{fig:gsefreq} shows the amplitude
variations of GDSE as a function of frequency for a number of lines of
sight. Each line of sight has a slightly different power law index
along the frequency direction as a result of the spatial variations in
the temperature spectral index. Furthermore, the brightness
temperature variation for one line of sight is not a single power law
but superposition of many power laws, due to the spectral index
variations both spatially and in the frequency direction. Note that $T_b$
is still a very smooth function of frequency.

\subsection{Emission from SNRs}

Supernova remnants are composed of expanding shells that have strong
magnetic fields which are able to produce cosmic rays. As the
particles escape the expanding shell, their energy decreases due to
synchrotron cooling and we detect them at radio frequencies.  The
majority of the Galactic SNRs are within the Galactic plane but their
distribution exponentially decreases with distance from the Galactic
plane, $z$, \citep[e.g.][]{caswell79, xu05}, that is, $N\sim {\rm
e}^{-z}$.  Moreover, due to the interaction of SNRs with the
interstellar medium their radio surface brightness $\Sigma$ decreases
with an increase of their diameter $D$ and with an increase of their
height $z$, \citep[e.g.][]{caswell79}, namely $\Sigma\sim D^{-3}{\rm
e}^{-z}$.

Our goal is to calculate the expected number of known SNRs within a
LOFAR-EoR observational window at high Galactic latitudes, using the
known number of observed radio SNRs from the \citet{green06} catalogue
and assuming that their distribution follows $N\sim {\rm e}^{-z}$. On
average, we obtain between one and two known SNRs in each
$5^\circ\times 5^\circ$ observational window. Given the extended
nature of the SNRs we include two of them in each window in order to
examine the influence of bright extended sources on the calibration
process and foreground removal.

The simulated SNRs assume a power law spectrum:
\begin{equation}\label{pls}
S_{\nu}=S_{0} \left( \frac{\nu}{\nu_{0}} \right)^{\alpha},
\end{equation}
where $S_{\nu}$ is the flux density of a SNR at frequency $\nu$,
$S_{0}$ is its value at normalization frequency $\nu_{0}$, and
$\alpha$ is the spectral index.

The simulated SNRs are placed randomly on the map and their angular
size, flux density and spectral index are arbitrary chosen from the
\citet{green06} catalogue. The SNRs are added on the map as disks with
uniform surface brightness.

Properties of the two SNRs included in our foreground simulations are
shown in Table~\ref{snr}.
\begin{table}
  \centering
  \caption{Angular size, flux density at $150~{\rm MHz}$ ($S_{150~{\rm
  MHz}}$), spectral index ($\alpha$) and position on the map of the
  two simulated supernova remnants. Values are calculated according to
  the data in \citet{green06} catalogue. }
  \begin{tabular}{@{}ccccc@{}}
    \hline & angular size & $S_{150~{\rm MHz}}$ & $\alpha$ & position
    on the map\\ &[arcmin]& [Jy] & & [arcmin,arcmin]\\ \hline
    SNR{\tiny I} & $14\times11$ & 7.91 & -0.65 & (254,53)\\ SNR{\tiny
    II} & $5\times6$ & 14.30 & -0.4 & (102,212)\\ \hline
  \end{tabular}\label{snr}
\end{table}

\subsection{Diffuse free-free emission}
The diffuse thermal (free-free) emission contributes only $\sim 1$ per
cent of the total foregrounds within the frequency range of the
LOFAR-EoR experiment \citep{shaver99}. It arises due to bremsstrahlung
radiation in very diffuse ionized gas, with a total emission measure
of about $5~{\rm pc~cm^{-6}}$ at high Galactic latitudes and
$T_{e}=8000~{\rm K}$ \citep{reynolds90}. This gas is optically thin at
frequencies above a few MHz, so its spectrum is well determined and
has a temperature spectral index of $\beta=-2.1$.
 
At high Galactic latitudes, H$\alpha$ and free-free emission of the
diffuse ionized gas are both proportional to the emission
measure. Therefore, the Galactic H$\alpha$ survey is generally used as
a tracer of the Galactic diffuse free-free emission \citep{smoot98}.
However, some groups also find significant correlation between
free-free emission and dust emission \citep{kogut96, deoliveira97}
which can also be used as another independent tracer of the Galactic
free-free emission.

In our simulations we followed \citet{tegmark00} and \citet{santos05}
who included the Galactic diffuse free-free emission as a separate
component of the Galactic foregrounds with an angular power spectrum
$C_{l}\sim l^{-3.0}$ and frequency $T_{b}\sim\nu^{-2.15}$. Despite its
small contribution to the foregrounds, the free-free emission is
important for two reasons. Firstly, the amplitude of its angular
fluctuations is much larger than that of the EoR signal. Secondly, and
more importantly, its spectral index along the frequency direction is
quite different from the other foreground components and could be
important in testing the algorithms for the EoR signal extraction.

To obtain the Galactic free-free emission maps we followed the same
procedure as for the Galactic synchrotron emission with the additional
simplification of fixing the power law index $\beta$ to $-2.15$ across
the map. $T_{b}$ is normalized according to the relation between
H$\alpha$ and free-free emission \citep[see review by][]{smoot98}
whereby the intensity of H$\alpha$ emssion, $I_{\alpha}$, is:
\begin{equation}\label{Ialpha}
  I_{\alpha}=0.36~{\rm R}\left(\frac{EM}{{\rm
  pc~cm^{-6}}}\right)\left(\frac{T_{e}}{10^{4}~{\rm
  K}}\right)^{-\gamma},
\end{equation}
where $EM$ is total emission measure and $T_{e}$ temperature. For
$T_{e}<2.6 \times 10^{4}~{\rm K}$ the value of $\gamma$ is
$0.9$. Combining Eq.~\ref{Ialpha} with the free-free equations in
\citet{smoot98}, one finds a relation between $I_{\alpha}$ and
brightness temperature of free-free emission, $T_{b}^{ff}$:
\begin{equation}\label{Tff}
  T_{b}^{ff} (30~{\rm GHz})=7~{\rm \mu K}\left(\frac{I_{\alpha}}{{\rm
  R}}\right).
\end{equation}
Using Eq.~\ref{Ialpha}\&\ref{Tff} together with $EM=5~{\rm
pc~cm^{-6}}$ and $T_{e}=8000~{\rm K}$, for high Galactic latitudes,
one gets $T_{b}^{ff}(30~{\rm GHz})=15.4{~\rm \mu K}$. Assuming a
frequency power law spectrum for $T_{b}^{ff}$ with index $-2.15$, one
obtains $T_{b}^{ff}=2.2~{\rm K}$ at 120~{\rm MHz}.

\begin{figure}
\centering
\vspace{-0.4cm}
\hspace{-0.5cm}
\includegraphics[width=.5\textwidth]{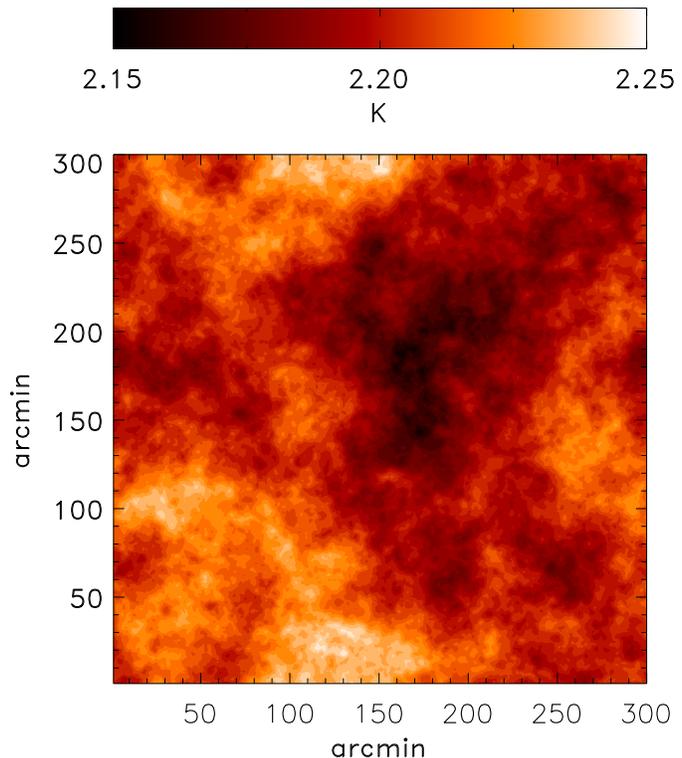}
\vspace{-1.0cm}
\caption{\emph{Simulated map of total intensity emission of Galactic
diffuse free--free emission with angular spectral index $\alpha=-3$
and frequency spectral index $\beta=-2.15$. The angular size of the
map is $5^{\circ}\times5^{\circ}$, with $\sim0.6'$ resolution. The
colour bar represents the brightness temperature $T_{b}$ of the
Galactic diffuse free--free emission in Kelvin at $120~{\rm MHz}$.}}
\label{fig:gffe}
\end{figure}

Fig.~\ref{fig:gffe} shows a simulated map of Galactic diffuse free-free
emission at $120~$MHz. The angular size of the map is $5^\circ\times
5^\circ$ on $512^{2}$ grid, with the mean brightness temperature of
$T_{b}=~{2.2 \rm K}$ and $\sigma={0.05 \rm K}$.

\section{Extragalactic foregrounds}
\label{sec:extragalactic}
\subsection{Radio galaxies}
At the frequency range of the LOFAR-EoR experiment, bright radio
sources are dominated by radio-loud galaxies, quasars and BL Lac
objects (an AGN class of objects). However, at sub-mJy flux densities
the contribution of late-type (star forming) galaxies, whose radio
synchrotron emission originates from supernovae rather than AGN,
becomes significant \citep{pradoni01, magliocchetti02, sadler02}

The bright extragalactic radio sources are normally divided into two
classes based on the relative physical position of their high and low
surface brightness area within the lobes. These two classes are called
FR{\small I} and FR{\small II} radio sources \citep{fanaroff74,
jackson99}.  Our simulations of radio galaxies are based on the tables
by \citet{jackson05} of extragalactic radio source counts at
151~{\rm MHz}. \citet{jackson05} has used $\Lambda$CDM based models to
calculate the evolution of the radio luminosity function of these
sources, from which was predicted the source distributions and their
number densities.

In obtaining these tables, \citet{jackson05} assumed that the radio
sky consists of three population types of radio sources: FR{\small I},
FR{\small II} and star forming galaxies. Moreover, it is assumed that
the local radio luminosity function of star forming galaxies can be
determined from the 2dFGRS-NVSS \citep{sadler02} galaxy sample at
1.4~{\rm GHz}. The parameterized number density and luminosity evolution
of star forming galaxies is adopted from \citet{haarsma00}. For the
local radio luminosity function of FR{\small I} and FR{\small II}
radio galaxies, \citet{jackson05} assumed that it can be determined by
exponential fitting of luminosity-dependent density evolution to the
observed source counts at $151~{\rm MHz}$ \citep{jackson99}. Given the
three evolving radio luminosity functions, \citet{jackson05} simulated
the sky at different frequencies by randomly positioning and
orientating each source on the sky. The intrinsic size is also
selected randomly, assuming redshift independence. The FR{\small I} and
FR{\small II} sources were modelled as double-lobe structures, and the
star forming galaxies as circular discs.

In our simulations of radio galaxies we adopt the three types of radio
sources from \citet{jackson05} and use the predicted source surface
densities per deg$^{2}$ for 10, 5, 2, 1 and 0.1~mJy flux density limit
in order to obtain the number of sources with certain flux density per
deg$^{2}$. However, and in contrast to the simulations by Jackson
where each source is randomly positioned, we introduce a angular
clustering of the sources. The clustering is motivated by the results
of \citet{dimatteo04} in which they showed that the contribution of
the angular clustering of extragalactic radio sources to the angular
fluctuations of the foregrounds, at scales $\gtrsim 1~{\rm arcmin}$,
is dominated by bright sources. Hence, in order to detect angular
fluctuations in the cosmological 21-cm emission, efficient source
removal $S \gtrsim 0.1 {\rm mJy}$ should be carried out.

For angular clustering of the radio galaxies we used the particularly
elegant procedure of Rayleigh-L{\`e}vy random walk proposed by
\citet{mandelbrot75, mandelbrot77}. Starting from any arbitrary
position, one places the next galaxy in a randomly chosen direction at
angular distance $\theta$, drawn from the distribution:
\begin{equation}\label{P(theta)}
  P(>\theta)= \left\{ \begin{array}{ll}
    (\theta / \theta_{0})^{\gamma} &{\rm for~} \theta \geq \theta_{0} \\
    1 &{\rm for~} \theta < \theta_{0},~\gamma >0 . \\
  \end{array} \right.
\end{equation}
This is repeated many times until the correlation function of the
distribution converges to the one desired. However, to compare the
introduced correlation with obervational results and set the right
values of $\gamma$ and $\theta_{0}$, the two point correlation
function needs to be calculated.

The two point correlation function, $w(\theta)$, of the radio galaxy
population is defined as the excess probability, over that expected
for a Poisson distribution, of finding a galaxy at an angular distance
$\theta$ from a given galaxy \citep[e.g.~][]{peebles80}:
\begin{equation}\label{prob}
\delta P=n[1+w(\theta)]\delta\Omega ,
\end{equation}
where $\delta P$ is probability, $n$ the mean surface density and
$\delta \Omega$ a surface area element.  Given the very large number
of galaxies that can be simulated, we adopted the simplest form for
estimating $w(\theta)$, defined as:
\begin{equation}\label{w(theta)}
w(\theta)=\frac{N_{\rm D}(\theta)}{N_{\rm R}(\theta)}-1 ,
\end{equation}
where $N_{\rm D}$ is the number of pairs of galaxies with separation
$\theta$ in the correlated sample of galaxies and $N_{\rm R}$ is the
number of pairs with the same separation $\theta$ but in a randomly
distributed uncorrelated sample of galaxies. The total number of
galaxies of the two samples must be the same.

Recent results on the angular clustering of radio sources in the NRAO
VLA Sky Survey (NVSS) and Faint Images of the Radio Sky at Twenty
centimetres (FIRST) \citep{overzier03} showed that the two point
correlation function is best fitted by a double power law
$w(\theta)={\cal{B}}\theta^{1-\gamma_{{\cal{B}}}}+{\cal{A}}\theta^{1-\gamma_{{\cal{A}}}}$
with slopes of $\gamma_{{\cal{B}}}=4.4$, $\gamma_{{\cal{A}}}=1.8$ and
amplitudes ${\cal{B}}=(1.5\pm0.2)\times 10^{-6}$,
${\cal{A}}=(1.0\pm0.2)\times 10^{-3}$. However, in this study we adopt
the simpler single power law correlation function with only two
parameters, $\gamma=1.8$ and ${\cal{A}}=0.002$. The reason for doing
this is that steeper power law component ($\gamma_{{\cal{B}}}$) of the
observed two point correlation function is dominant on angular scales
smaller than those resolved by the LOFAR-EoR experiment. Note that
$\gamma$ in the two point correlation function is the same one as in
Eq.~\ref{prob}, while ${\cal{A}}=(1/\theta_{0})^{\gamma}$.

\begin{figure}
\centering
\vspace{-0.5cm}
\hspace{-0.5cm}
\includegraphics[width=.5\textwidth]{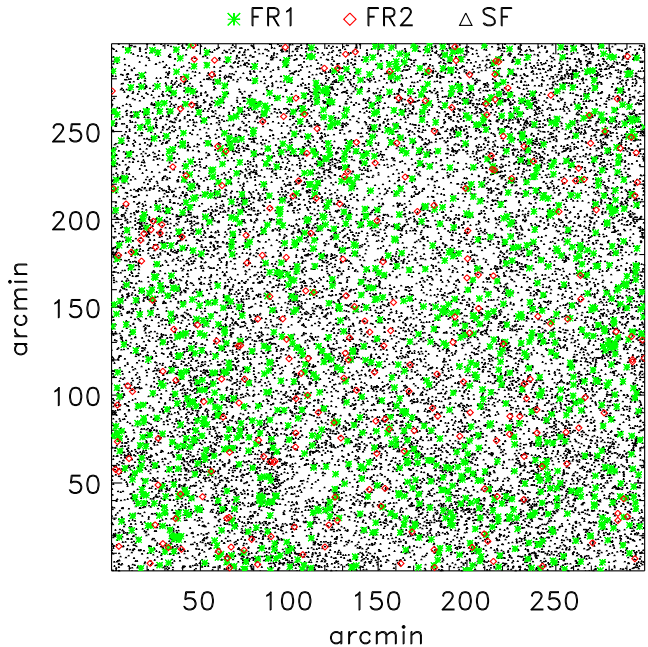}
\vspace{-1.0cm}
\caption{\emph{Simulated 2D angular clustering of radio galaxies (
$92$ per cent - star forming (SF), $7$ per cent - FR{\scriptsize I}
and $1$ per cent - FR{\scriptsize II} radio galaxies) with angular
correlation function of the form $w(\theta)=0.002\theta^{-0.8}$
(${\cal{A}}=0.002$, $\gamma=1.8$). The flux density distribution of
the galaxies is shown in the Table~\ref{rgtable}. The angular size of
the map is $5^{\circ}\times5^{\circ}$.}}
\label{fig:rgalax}
\end{figure}

Fig.~\ref{fig:rgalax} shows a simulated map of radio galaxies with
angular power law distribution. All sources are point like as the
angular resolution of the LOFAR-EoR project will not be sufficient to
resolve most of them. The angular extent of the map is
$5^{\circ}\times5^{\circ}$. On the map there are in total 20690 radio
galaxies: $92$ per cent SF radio galaxies, $7$ per cent FR{\small I}
and $1$ per cent FR{\small II} radio galaxies. The flux density
distribution at $150~{\rm MHz}$ of the simulated radio galaxies are
shown in the Table~\ref{rgtable}, while the two point correlation
fuction is $w(\theta)=0.002\theta^{-0.8}$ (${\cal{A}}=0.002$,
$\gamma=1.8$).  The simulated radio galaxies assume a power law
spectrum (see Eq.~\ref{pls}) with spectral index -0.7
\citep[e.g.][]{jackson05}.

\begin{table}
  \centering
  \caption{The flux density distribution of the simulated radio
galaxies on the $5^{\circ}\times5^{\circ}$ map
(Fig.~\ref{fig:rgalax}). The corresponding frequency is $150~{\rm
MHz}$.}
  \begin{tabular}{@{}cccccc@{}}
    \hline &\multicolumn{5}{c}{number of sources with flux density
   limit} \\ &10 mJy& 5 mJy & 2 mJy & 1 mJy & 100 $\mu$Jy \\ \hline
   FR{\scriptsize I}&20&55&122&177&1001\\ FR{\scriptsize
   II}&25&39&48&54&89\\ SF&4&38&210&419&18379\\ \hline
  \end{tabular}\label{rgtable}
\end{table}

\subsection{Radio clusters}
Galaxy clusters as radio sources are classified into cluster radio
haloes and cluster radio relics.  The former are morphologically regular
diffuse sources, typically centred inside the cluster and mostly
unpolarized, whereas the radio relics are typically irregular, located
at the periphery of the cluster and consist mostly of polarized radio
diffuse sources. Both types of cluster radio source have steep
frequency spectra with $\beta \sim -3$ \citep[see e.g.][for
review]{feretti02}.

The emission in radio haloes is due to synchrotron radiation by
relativistic electrons with energies of $\sim 10$~GeV in $\mu$G
magnetic fields. The distribution of the radio haloes seems to follow
closely the large scale distribution of the free-free driven X-ray
emission of clusters \citep{govoni01}. This association is also
supported by a strong correlation between the radio halo luminosity
and the host cluster X-ray luminosity \citep[e.g.][]{ensslin02}.
However, not all clusters host radio haloes. Statistically, it is found
that roughly 30--40 per cent of galaxy clusters with X-ray luminosity
$L_{\rm X}\ge 10^{45}~{\rm erg~s^{-1}}$ do host radio haloes.

In our simulations of extragalactic foregrounds maps, as a starting
point for simulating radio clusters, we used the $\Lambda$CDM deep
wedge cluster catalogue of The Hubble Volume Project\footnote{The
cluster catalogue is part of simulations carried out by the Virgo
Supercomputing Consortium using computers based at the Computing
Centre of the Max-Planck Society in Garching and at the Edinburgh
parallel Computing Centre.  The data are publicly available at
http://www.mpa-garching.mpg.de/galform/virgo/hubble.}. The catalogue
was obtained from an N-body simulation with one billion dark matter
particles. This catalogue provides a list of clusters up to redshift
$z \le 4.37$ with angular coverage of 100~deg$^{2}$
\citep{colberg00,jenkins01,evrard02}.

In order to translate the cluster mass into X-ray luminosity $L_{\rm
X}$ and then into radio luminosity $L_{\rm r}$, we used
the empirical mass--X-ray luminosity relation of
\citet{reiprich02}:
\begin{equation}\label{MXRL}
L_{\rm X}=a_{\rm X}10^{45}h^{-2}_{50}~{\rm
erg~s^{-1}}\left(\frac{M_{\rm R\&B}}
{10^{15}h^{-1}_{50}M_{\sun}}\right)^{b_{\rm X}}
\end{equation}
and the X-ray--radio luminosity relation of \citet{ensslin02}:
\begin{equation}\label{XRRL}
L_{\rm r, 1.4~GHz}=a_{\rm r}10^{24}h^{-2}_{50}~{\rm
W~Hz^{-1}}\left(\frac{L_{\rm X}} {10^{45}h^{-2}_{50}~{\rm
erg~s^{-1}}}\right)^{b_{\rm r}}
\end{equation}
where $a_{\rm X}=0.449$, $b_{\rm X}=1.9$, $a_{\rm r}=2.78$ and $b_{\rm
r}=1.94$ \citep{ensslin02}. It is important to note that
Eq.~\ref{XRRL} is derived for the radio frequency at 1.4~GHz. Note
also that in Eq.~\ref{MXRL} the mass, $M_{\rm R\& B}$, is related to
the cluster real mass, $M$, by $M_{\rm R\& B}\approx M
\Omega^{1/2}_{m}$.

Since $L_{\rm r}$ is derived for 1.4~GHz, we extrapolate the
luminosity of each cluster to lower frequencies, relevant to the
LOFAR-EoR experiment, according to:
\begin{equation}\label{LrLOFAR}
L_{r}(\nu)=L_{\rm r,1.4~GHz}\left(\frac{\nu
[MHz]}{1400}\right)^{\alpha},
\end{equation}
where $\alpha=-1.2$ \citep{kempner04}.

The angular size of the radio clusters was estimated from their
physical radius and redshift. For the physical radius we used the virial
radius $r_{vir}$ calculated from the cluster mass according to \citep{bryan98},
\begin{equation}\label{MvirR}
M=4\pi r^{3}_{vir} \rho_{crit} \Delta_{c}/3
\end{equation}
where $\Delta_{c}$ is the mean density and $\rho_{crit}$ is the
critical density at redshift $z$.

In order to obtain the maps of radio clusters at the angular and
frequency ranges desired we first randomly choose 30 per cent of the
catalogue's clusters (note that observations show that only 30--40 per
cent of clusters have radio properties).  The cluster masses are then
used to estimate the radio luminosity of each cluster
(Eq.~\ref{MXRL},~\ref{XRRL}~\&~\ref{LrLOFAR}) and its virial radius
(Eq.~\ref{MvirR}). Finally, the cluster is projected onto the
simulated map according to its survey coordinates in the Hubble Volume
simulation, its redshift and estimated size. Note that the radio
clsuters are added on the map as disks with uniform surface
brightness.

\begin{figure}
\centering
\hspace{-0.5cm}
\includegraphics[width=.5\textwidth]{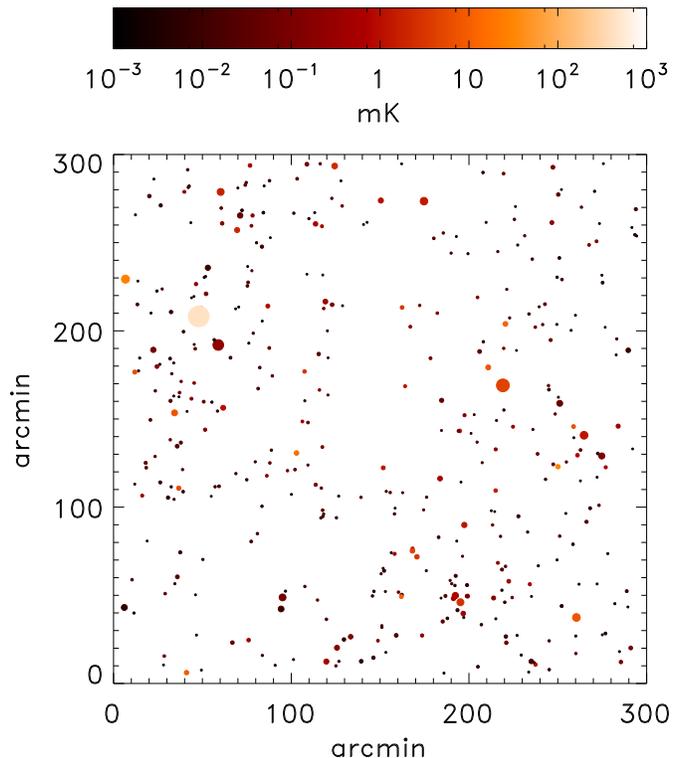}
\vspace{-1.0cm}
\caption{\emph{Simulated 120~{\rm MHz} map of radio clusters of
galaxies. The angular size of the map is
$5^{\circ}\times5^{\circ}$. Note that the colour bar represents the
brightness temperature of the clusters in logarithmic units and that
the size of each cluster has been scaled by a factor of 10 for visual
clarity. We also assume uniform surface brightness of the clusters.}}
\label{fig:rclus}
\end{figure}

Fig.~\ref{fig:rclus} shows a $5^{\circ}\times5^{\circ}$ map of radio
clusters simulated at $120~{\rm MHz}$. The colour bar represents the
brightness temperature of the clusters in logarithmic units. The size
of each cluster has been scaled by a factor of 10 for visual clarity.

\section{Polarization}
\label{sec:polarization}

The need for understanding the polarization response stems from two
main factors. One is the geometry of the LOFAR configuration, and the
other is the cross-talk contamination between the two dipoles of a
LOFAR antenna. In order to detect the EoR signal, which has at best a
signal to noise ratio (SNR) of $\approx 0.2$ per beam on most angular
scales -- assuming 400 hours of observation under the configuration
specified in Section \ref{sec:inst} -- we need to fully understand the
response of the LOFAR system in total intensity and polarization. As
discussed before, this is vital for us in order to be able to span a
dynamic range of 4--5 orders of magnitude.

Since the LOFAR antennae are fixed to the ground, the sources are
tracked only by beam-forming and not by steering the antennae as in
traditional radio astronomy. This implies that, depending on the
position of the source on the sky, only a certain projection of the
two dipoles is visible. This projection changes as the source is
tracked over time. Therefore, at most times the sources in the sky see
different projections of each of these dipoles. Now, if the
sources/foregrounds are polarized, we immediately see that the power
output from the pair of dipoles (which is the sum of the two polarized
components) will vary dramatically. This has to be taken in account
almost exactly during the inversion and calibration processes in order
to achieve the desired dynamic range.

On the other hand, a leakage in the electronics can cause the power
that is supposed to go through one of the dipoles to show up in the
other (referred to as cross-talk). Although the cross-talk is small
compared to the effect due to projection, we need to take it into
account to eliminate any systematics.

In this paper we show a first order simulation of the Galactic diffuse
polarized emission and defer a more advanced discussion on the topic
to a future paper.

There are several polarization surveys of the Galactic synchrotron
emission between $327~$MHz and $2.7~$GHz \citep[see
review,][]{reich06}.  The most recent one was done with the
Westerbork telescope at $327~$MHz, with arc minute angular resolution
\citep{wieringa93, haverkorn00, haverkorn02, haverkorn03}. Its low
frequency maps reveal a large amount of unusually shaped polarized
small-scale structures, which have no counterpart in the total
intensity. These structures are normally attributed to the coexistence
of magnetic fields and thermal gas in the interstellar medium, which
produce Faraday rotation at each line of sight.

The Faraday rotation depends on the observing frequency and rotation
measure (RM) of the structure and it is defined along the line of
sight. In order to measure Faraday rotation, observations at two or
more frequencies are required. However, full understanding of the
observed results could be quite difficult due to the possibility of
multiple Faraday rotation layers (screens) along the line of sight and
depolarization effects. \citet{brentjens05} introduced a new method
(Faraday Rotation Measure Synthesis) that is able to cope with
multiple screens and analyzes the contribution of each of these
screens separately.

The Galactic diffuse synchrotron emission is linearly polarized and
its polarized intensity $I_{P}$ can be expressed in terms of Stokes
parameters $U$ and $Q$:
\begin{equation}
I_{P}=\sqrt{U^{2}+Q^{2}}
\end{equation}
\begin{equation}
\theta=\frac{1}{2}\arctan{\frac{U}{Q}}
\end{equation}
where $\theta$ is the polarization angle.

In order to simulate the polarization maps of the Galactic synchrotron
emission at low radio frequencies, we use a simple model of the
Galactic synchrotron polarization at high frequencies
\citep{giardino02} in combination with Faraday screens that are
introduced to account for the effects of rotation and depolarization
at low frequencies.  Note that the assumption about the correlation
between the polarized and total intensity going into the Galactic
synchrotron polarization model at high frequencies is not valid at low
frequencies, since the observations mentioned above show polarized
structures that have no counterpart in the total intensity. However,
this assumption can be acceptable for a first estimate. Note that the
correlation assumption should not be used in the analysis of the
redshifted 21 cm data as it is mostly invalid and can lead to wrong
interpretations. In the future, we will improve on the model itself
and use results from real polarization data obtained by the
LFFE\footnote{LFFE (Low Frequency Front End) are receivers at the
Westerbork Synthesis Radio Telescope (WSRT) and cover the frequency
range from 115 to 170 MHz}.

The basic assumptions of the \citet{giardino02} model of Galactic
synchrotron polarization at high frequencies are:
\begin{enumerate}
\item The polarized component of Galactic synchrotron emission is
proportional to the unpolarized intensity, which in terms of
brightness temperature $T_{b}$ is:
  \begin{equation} \label{QU}
    Q=fT_{b}\cos(2\theta)
  \end{equation}
  \[U=fT_{b}\sin(2\theta)\]
  where $f$ is the fraction of polarized emission (or polarization
  degree) and $\theta$ is the polarization angle;
\item The fraction of polarized radiation $f$ is related to the
  temperature spectral index $\beta$ \citep{cortiglioni95}:
  \begin{equation}\label{f}
   f=\frac{3\beta-3}{3\beta-1}
  \end{equation}
\item The polarization angle $\theta$ is given by:
  \begin{equation}\label{theta}
    \theta=\frac{1}{2}\arctan(x/n,y/n)
  \end{equation}
where $x,y$ are 2D random Gaussian fields with the mean zero
and characterized by a power law spectrum, while
$n=\sqrt{x^2+y^2}$. The power law spectral index is $\alpha=-1.7$ and
its value is driven by observations, e.g. the Parkes 2.4 GHz
polarimetric survey \citep{duncan95}.

\end{enumerate}

The Faraday screens are modelled as 2D fields of rotation angles
$\Delta \theta$ defined by \citep{rybicki86}:
\begin{equation}\label{Dtheta}
\Delta \theta = RM \lambda^{2}
\end{equation}
where $\lambda$ is the wavelength of radiation and $RM$ is the
rotation measure modelled as a 2D Gaussian random field (GRF) with a
power law spectrum of spectral index $\alpha$. Note that for the
demonstrative purpose of this simulation we introduced only two
Faraday screens, with mean zero and standard deviation 0.3, and
arbitrarily set the value of $\alpha$ to -2.

Therefore, in order to generate polarization maps of Galactic
synchrotron emission at given frequencies, first we take the GDSE maps
of total intensity ($T$) and temperature spectral index ($\beta$) from
Sec.~\ref{gdse} and calculate the fraction of polarized radiation
according to Eq.~\ref{f}. Then, using Eqs.~\ref{theta} \& \ref{Dtheta} we
obtain polarization angle $\theta$ and Faraday rotation angle $\Delta
\theta$. Finally, we use Eq.~\ref{QU} to get polarization maps Q
and U. The angle in Eq.~\ref{QU} is the sum of $\Delta \theta$ over
all Faraday screens and $\theta$.

\begin{figure}
\centering
\hspace{-0.5cm}
\includegraphics[width=.5\textwidth]{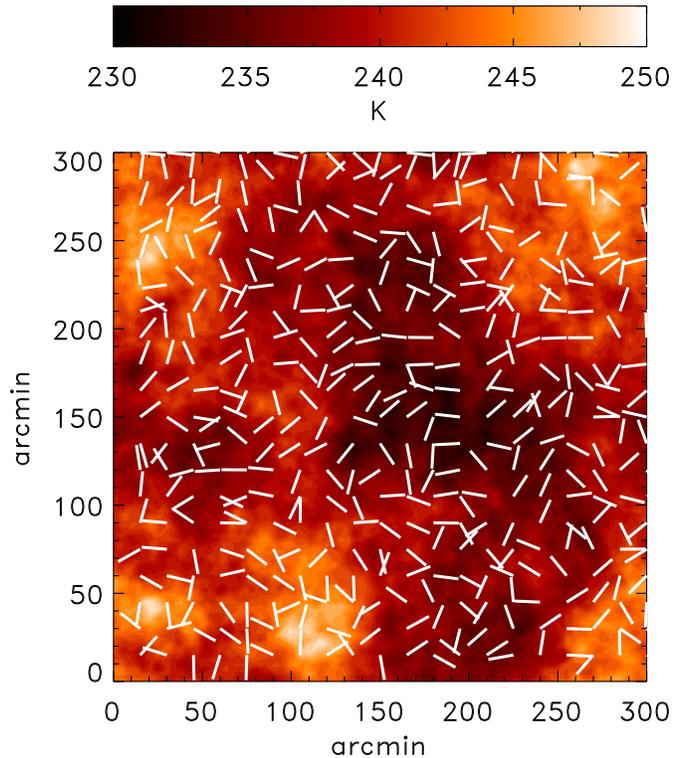}
\vspace{-1.0cm}
\caption{\emph{Simulated $120~{\rm MHz}$ map of polarized intensity
($I_{p}$) of diffuse Galactic synchrotron emission, and polarization
angle (white lines). The angular size of the map is
$5^{\circ}\times5^{\circ}$, with $\sim0.6'$ resolution.}}
\label{fig:Iplot}
\end{figure}

\begin{figure}
\centering
\hspace{-0.5cm}
\includegraphics[width=.5\textwidth]{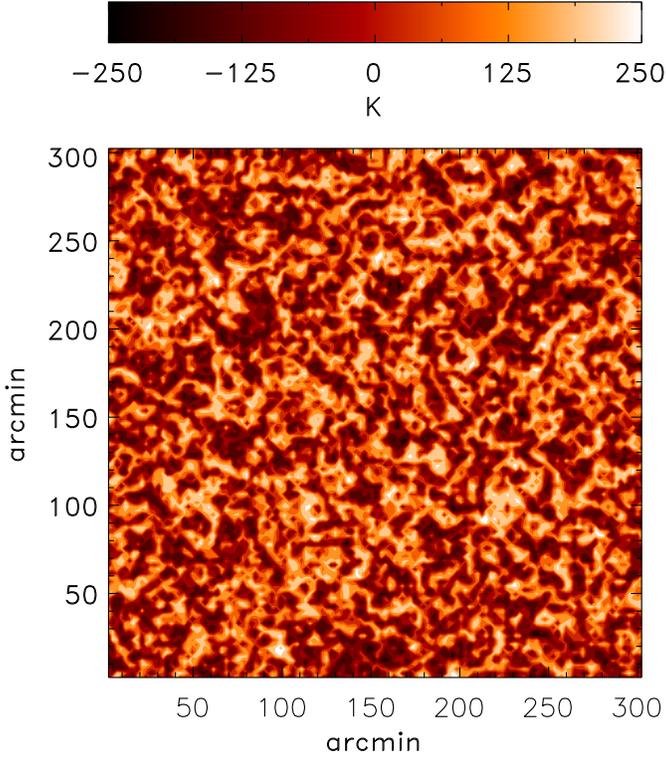}
\vspace{-1.0cm}
\caption{\emph{Simulated $120~{\rm MHz}$ Stokes Q map of polarized
diffuse Galactic synchrotron emission (DGSE). The angular size of the
map is $5^{\circ}\times5^{\circ}$, with $\sim0.6'$
resolution. Simulated Stokes U map of polarized GDSE looks very
similar to Q map.}}
\label{fig:Qplot}
\end{figure}

Fig.~\ref{fig:Iplot} shows the simulated 120~{\rm MHz} map of
polarized intensity ($I_{p}$) of diffuse Galactic synchrotron
emission. The polarization angles are shown as white lines. The Stokes
Q map of simulated Galactic polarized emission is shown in
Fig.~\ref{fig:Qplot}. The related Stokes U map looks very similar to
the Q map.

\section{Instrumental effects}
\label{sec:inst}
\begin{figure}
\centering
\hspace{-0.5cm}
\includegraphics[width=.5\textwidth]{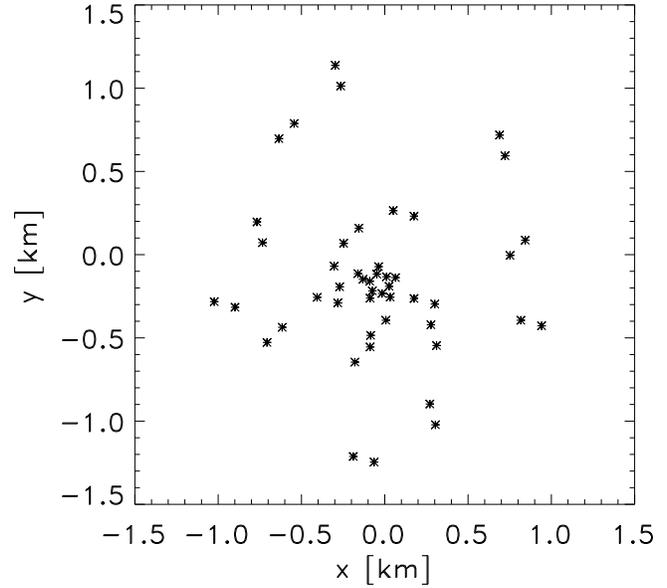}
\vspace{-0.5cm}
\caption{\emph{Position of 48 substations (24 stations) of the LOFAR
    core used for simulations of instrumental effects. Note that each
    substation consists of 24 tiles, with each tile having 4$\times$4
    crossed dipoles.}}
\label{lofar}
\end{figure}

In this section we give a basic overview of the simulations of LOFAR
antenna response and show how the foreground maps are seen by
LOFAR. More detailed discussion on the LOFAR response and data model
for the LOFAR-EoR experiment will appear in \citet[][in
prep.]{panos08}. For the LOFAR-EoR experiment we plan to use the LOFAR
core, which will consist of approximately 25 stations. However, in
this paper we set the number of LOFAR core stations to 24.  Each
station is further split into two substations which are separated by a
few tens of metres (see Fig.~\ref{lofar}). Each substation consists of
24 tiles, with each tile having 4$\times$4 crossed dipoles. For our
goals we assume that each of the forty-eight substations is a circular
array with a radius of thirty-five metres. The stations are
distributed in a randomized spiral layout and span a baseline coverage
from 40 to 2000 metres. The total effective collecting area for the
LOFAR-EoR experiment is $\sim0.07~{\rm km^2}$ at $150~{\rm MHz}$. The
instantaneous bandwidth of the LOFAR telescope is $32~{\rm MHz}$ and
the aim for the LOFAR-EoR experiment is to observe in the frequency
range between 115--180$~{\rm MHz}$, which is twice the instantaneous
bandwidth. To overcome this, multiplexing in time has to be used
\citep[for more details see,][]{PP07}. For the purpose of this paper
we ignore this complication and assume 400 hours of integration time
for the hole frequency range.  The LOFAR specifications are not final
yet and they might slightly change in the near future. A detailed
description of the telescope together with the data model will be
provided in forthcoming papers.

In order to compute the true underlying visibilities, we make some
simplifying assumptions. We assume that the narrow bandwidth condition
holds and that the image plane effects have been calibrated to a
satisfactory level. This includes station complex gain calibration, a
stable primary beam, and adequate compensation for the ionospheric
effects, such that the ionospheric phase introduced during the
propagation of electromagnetic waves in the ionosphere and the
ionospheric Faraday rotation are corrected for.  For an
interferometer, the measured spatial correlation of the electric
fields between two antennae is called `visibility' and is
approximately given by \citet{taylor99}:

\begin{eqnarray}\label{vis}
    V_{f}(u,v)=\int A(l,m)I_f(l,m) \nonumber {\rm e}^{i(ul+vm)}{\rm
    d}l{\rm d}m
\end{eqnarray}
where $A$ is the primary beam, $I_{f}$ is the intensity map corresponding
to frequency $f$, $(u,v)$ are the coordinates, as seen from the
source, of the tracks followed by an interferometer as the Earth
rotates, and $(l,m)$ are the direction cosines.

We further treat each pixel of the map as a point source with the
intensity corresponding to intensity of the pixel. Note that the
equation above takes into account the sky curvature. The visibilities
are sampled for all substation pairs and also at different
pair positions, as the Earth rotates.

We calculate the Fourier transform of the foreground model for each
frequency in the above range. For every baseline and frequency, the uv
tracks sample different scales of the Fourier transform of the
sky at that frequency. Thus, the sampling function $S$ becomes 

\begin{equation}
S(u,v)= \sum\limits_k {\delta \left( {u - u_k } \right)} \delta \left( {v-
v_k } \right),
\end{equation}
where the summation is carried over all the pixels $k$.

We compute those tracks for each interferometer pair for 4 hours of
synthesis with an averaging interval of $100~{\rm s}$ and we then grid
them on a regular grid in the uv plane. The maximum baseline assumed
for the LOFAR core is $2~{\rm km}$ and the station diameter is
$35~{\rm m}$, the number of independent elements in the uv plane is
$\approx 60^2$. If the uv plane is oversampled by a factor of four,
this yields $256^2$ pixels \footnote{This is the closest power of two
to match the number of sampled elements. By doing this one can benefit
from the speed of the FFT.}  in the uv plane of $\approx 60~\rm {m^2}$
in size.  After counting how many track points fall within each grid
cell, we end up with a matrix representing the naturally weighted
sampling function in the uv plane. By multiplying this sampling matrix
with the Fourier transform of our model sky we get the visibilities on
that grid with appropriate weights. This procedure is done for each
baseline pair.

\begin{equation}
V_f \left( {u,v} \right) = S \cdot \mathcal{F}I
\end{equation}
where $\mathcal{F}I$ is the Fourier transform of the input image and
$S$ is the sampling function.

\begin{figure*}
\centering
\vspace{-0.0cm}
\hspace{-0.5cm}
\includegraphics[width=.45\textwidth, angle=0]{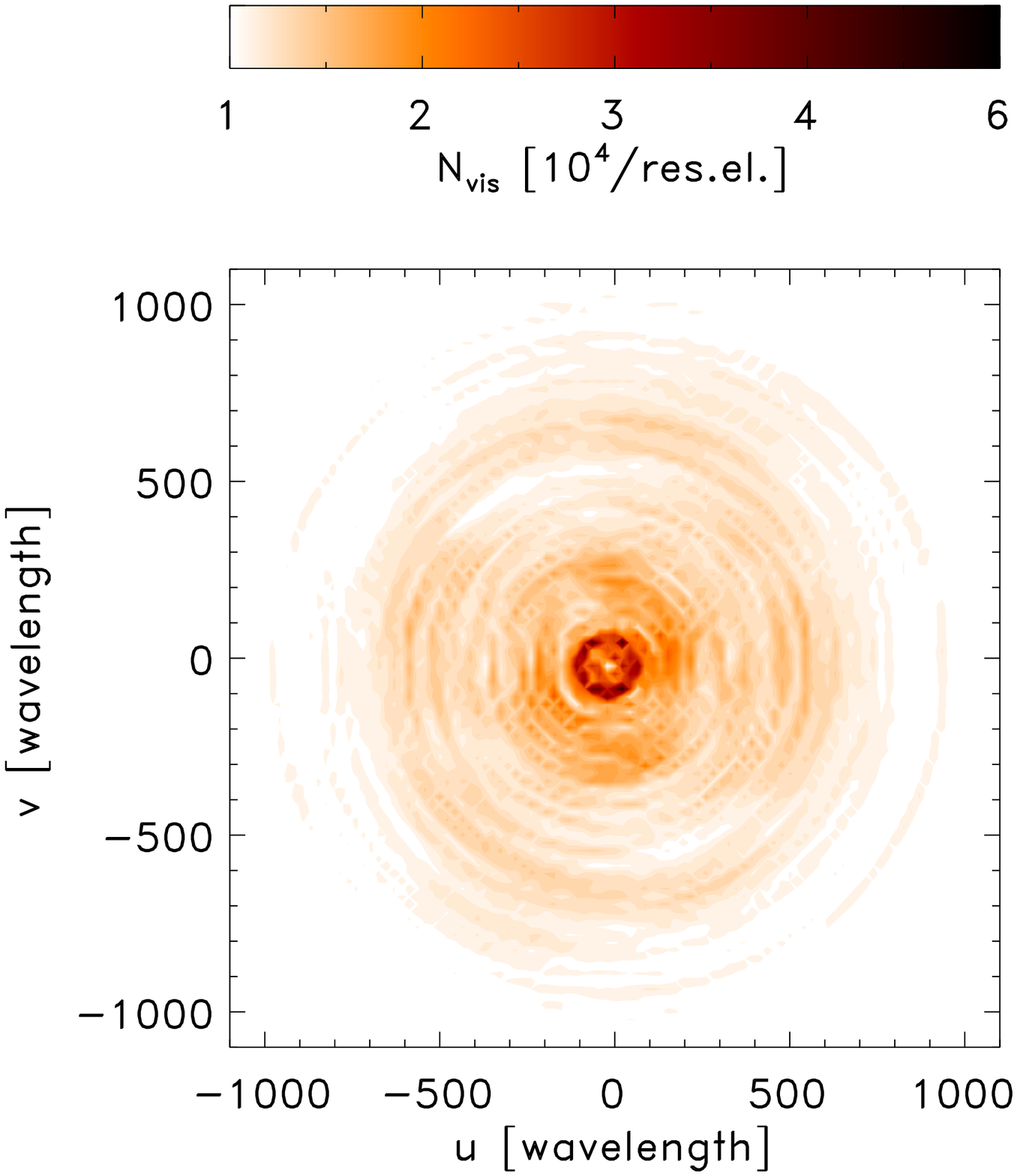}
\hspace{0.5cm}
\includegraphics[width=.45\textwidth, angle=0]{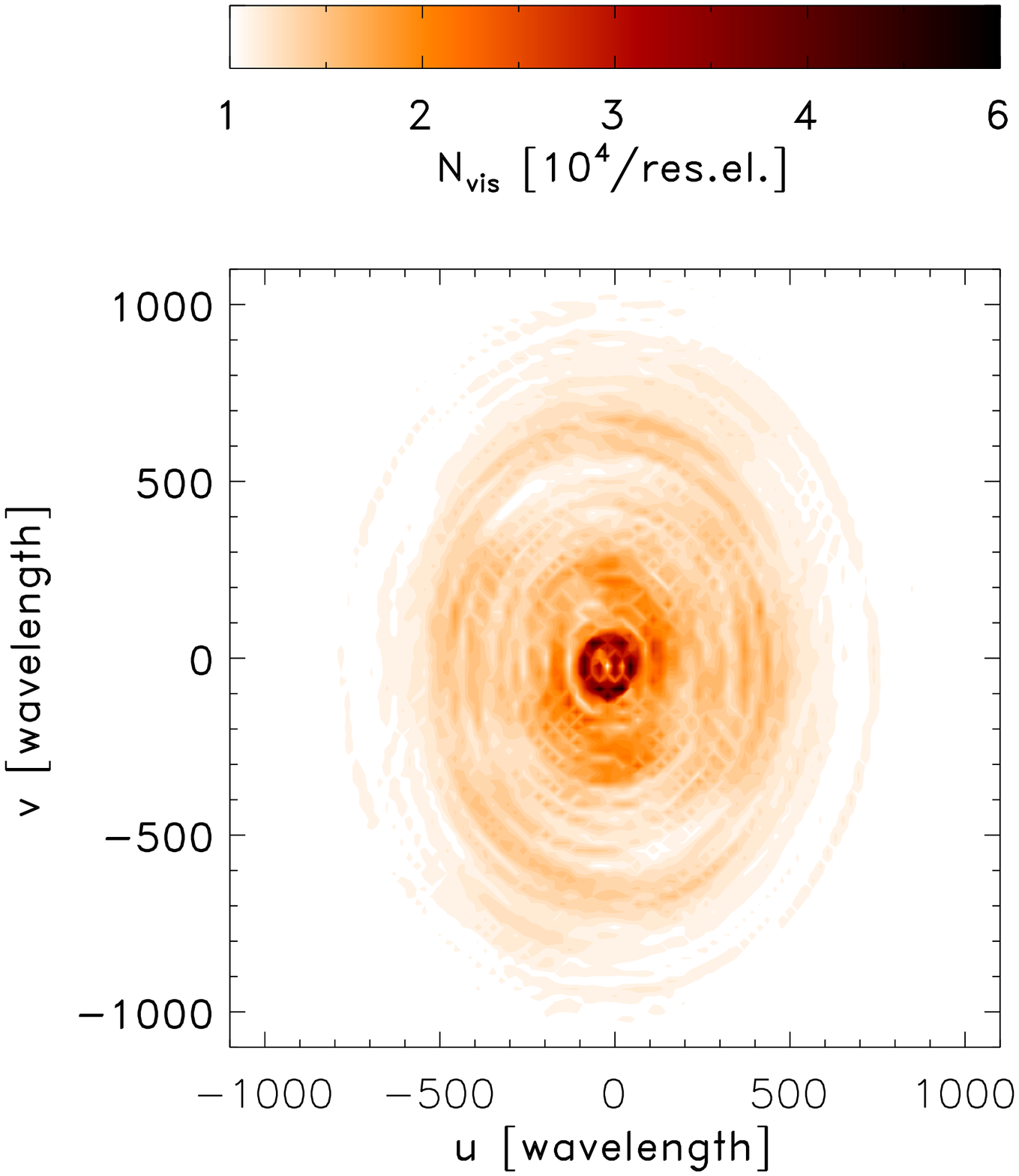}
\vspace{-0.5cm}
\caption{\emph{The expected LOFAR visibility densities per resolution
element at $150~{\rm MHz}$ for $400~h$ of total integration time
($100\times4~\rm{h~night^{-1}}$) with averaging time of $100~\rm{s}$
and for observing declinations $\delta = 90^{\circ}$ (left panel) and
$\delta = 52^{\circ}$ (right panel).}}
\label{fig:uvdens}
\end{figure*}

The LOFAR visibility densities per resolution element at 150~{\rm
MHz}, for the LOFAR-EoR experiment, are shown in
Fig.~\ref{fig:uvdens}. The total integration time is 400 hours
($100\times4~\rm{h~night^{-1}}$) with averaging time of $100~\rm{s}$
and observing declinations $\delta = 90^{\circ}$ (left panel) and
$\delta = 52^{\circ}$ (right panel).

\begin{figure*}
\centering
\hspace{-0.5cm}
\includegraphics[width=1.\textwidth]{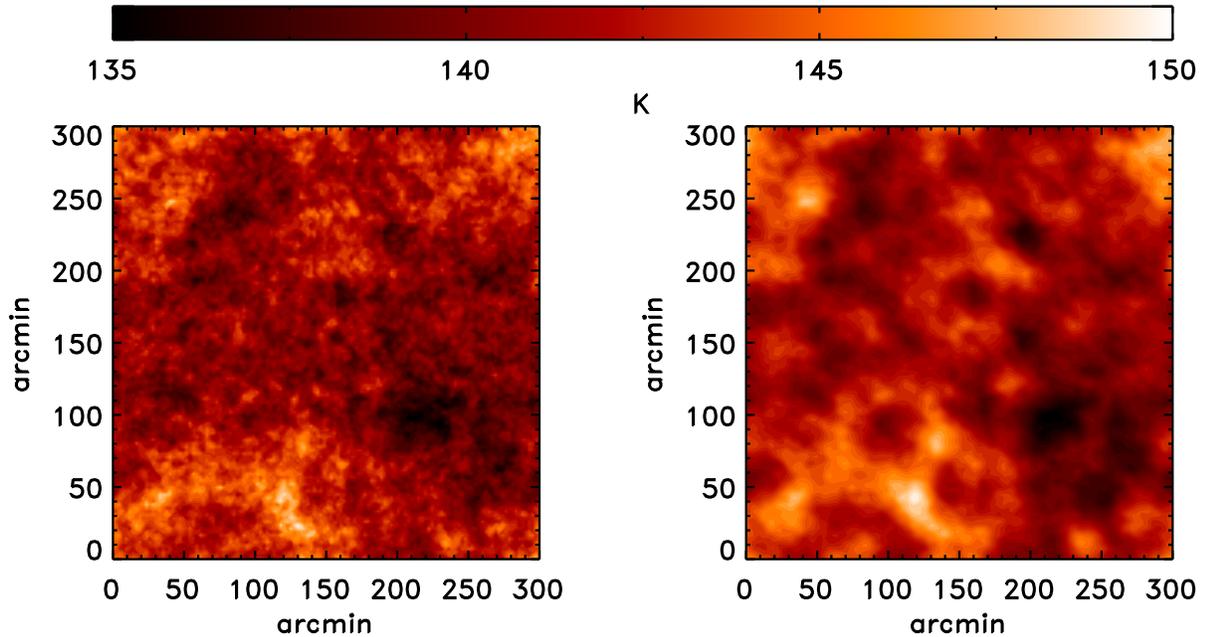}
\vspace{-0.0cm}
\caption{\emph{Total intensity map of the simulated diffuse components
of the foregrounds (`original' map with no interferometric effects and
noise) and its corresponding `dirty' map after $400~h$ of total
integration time with averaging of $100~{\rm s}$ at $150~{\rm MHz}$.}}
\label{fig:dmmap}
\end{figure*}

The inverse Fourier transform of the sampled visibilities is called
the `dirty' map. It is actually the sky map convolved with the Fourier
transform of the sampling function, which is called the `dirty' beam
or the `PSF'. This is a simple-minded approach to estimating the sky
brightness as it uses linear operations. The approximation of the
underlying brightness with the `dirty' map is not always satisfactory,
as side lobes from bright features will obscure fainter ones. In cases
of low signal to noise, however -- such as during the observation of the
redshifted 21-cm transition line of HI -- one might choose not to proceed
further than this first approximation. To go beyond that we need extra
information like the positivity of the intensity and compact
support. The discussion of such issues is beyond the scope of this
paper. This incomplete sampling of the uv plane also means that we do
not measure the complete power at all scales, due to the holes in the
uv coverage and its finite extent.

An example of a `dirty' map of the diffuse components in the
foregrounds is shown in the Fig.~\ref{fig:dmmap}, together with the
`original' simulated foreground map with no interferometric effects
and noise. The corresponding total integration time is 400 hours, with
an averaging time of $100~{\rm s}$ at $150~{\rm MHz}$. Note that the
`dirty' maps are generated without the inclusion of noise.

The ultimate sensitivity of a receiving system is determined
principally by the system noise. The discussion of the noise
properties of a complex receiving system like LOFAR can be lengthy, so
we concentrate for our purposes on some basic principles. The
theoretical rms noise level in terms of flux density on the final
image is given by

\begin{equation}
\sigma _{noise} = \frac{1}{{\eta _s }} \times \frac{{SEFD}}{{\sqrt {N
\times (N - 1) \times \Delta \nu \times t_{{\mathop{\rm int}} } } }}
\end{equation}
where $\eta _s$ is the system efficiency that accounts for electronic,
digital losses, N is the number of substations, $\Delta \nu$ is the
frequency bandwidth and $t_{{\mathop{\rm int}} }$ is the total
integration time. SEFD is the System Equivalent Flux Density in ${\rm
Jy}$. The system noise we assume has two contributions. The first
comes from the sky and is frequency dependent ($\approx \nu^{-2.55}$)
and the second from receivers. 

For the LOFAR core the SEFD will be around $1000~{\rm Jy}$ at
$150~{\rm MHz}$, depending on the final design \citep{PP07}. This
means that we can reach a sensitivity of $520~{\rm mK}$ at $150~{\rm
MHz}$ with $1~{\rm MHz}$ bandwidth in one night of observations. In
order to calculate the SEFD we use the following system temperature
($T_{sys}$) scaling relation as function of frequency ($\nu$):
$T_{sys}=140+60(\nu/300~\rm{MHz})^{-2.55}$. Accumulating data from a
hundred nights of observations brings the sensitivity down to $52~{\rm
mK}$. We further assume that the distribution of noise over the map at
one frequency is Gaussian. The noise contribution to each pixel in a
map is drawn independently from a Gaussian distribution. The EoR
signal is extracted from two different scenarios. The first scenario
involves the extraction of the signal from the `original' maps --
simulated maps that are not convolved with the dirty beam -- after
adding the noise directly to the `original' maps.  In the other
scenario, the EoR signal is extracted from `dirty' maps to which we do
not add noise but convolve the `original' maps of the EoR signal plus
the foregrounds with a simplified dirty beam.

As the uv coverage scales linearly with frequency, one has to be
careful in using the `dirty' maps for extraction. This is because a
pixel sampled at a given frequency need not be sampled at a later
frequency. Since the analysis performed in this paper involves data
across the frequency domain, we need a good uv coverage. If the uv
sampling functions are not scaled accordingly, we will introduce
additional difficulties arising from the mixing of spatial scales. To
overcome spatial scale mixing, one of the strategies in the data
analysis is to use only the uv points that are present at all
frequencies. In other words one can construct the uv plane mask that
only contains the uv points that are sampled at every frequency. This
step of course results in substantial data loss.

The uv coverage for the LOFAR-EoR experiment changes in scale by
$\sim40\%$ between the frequency range of observation
($115$--$180~{\rm MHz}$). By choosing only those uv points that are
present at all frequencies, $\sim 5\%$ of the total data is lost in
the frequency range specified above. Since with decreasing bandwidth
of observation the amount of data lost decreases, one of the
strategies could be to observe in windows of smaller
bandwidth. However the observational strategy of the LOFAR-EoR
experiment is not yet finalized and will be discussed in detail in
upcoming papers of the project. A detail discussion on the scaling of
the uv coverage with frequency and its influence on the number of
discarded baselines and the amount of data loss will be discussed in
\citep[][in prep.]{panos08}.

In the following section we will show our ability to statistically
detect the EoR signal from the `original' maps that include realistic
level for the noise and from `dirty' maps that do not include the
noise but are sampled with the uv mask that contains only uv points
present at each frequency. In both cases the statistical detection of
the signal is done on total intensity maps only. Moreover, perfect
calibration is assumed and any other systematics that might influence
the data are ignored. Those issues will be addressed in a follow-up
paper.

\section{Detection of the EoR signal from the FG} 
\label{sec:extraction}
This section presents the results on the statistical detection of the
EoR signal from `original' and `dirty' LOFAR-EoR data maps that
include the cosmological 21cm signal, diffuse components of the
foregrounds and realistic noise. By `original' maps we mean maps
before inversion or in other words maps with no calibration errors or
interferometric effects. `Dirty' maps include only simplified uv
coverage effect as an interferometric effect, but no calibration
errors.

By using only diffuse components of the foregrounds (Galactic diffuse
synchrotron and free-free emission and integrated emission from
unresolved extragalactic sources) we assume that all resolved discrete
and extended sources have been successfully removed from the observed
maps, without any subtraction residuals. Also note that our analysis
is done on total intensity maps only. The polarized case will be
considered in the follow-up paper.

The foreground and noise maps are simulated in the frequency range
between $115~{\rm MHz}$ and $178.5~{\rm MHz}$ in steps of $0.5~{\rm
MHz}$. The original maps simulated for a $5^{\circ}\times5^{\circ}$
field on a $512^{2}$ grid are re-binned to a $128^{2}$ grid, so that
each pixel corresponds to $2.3'$ which is the resolution attained by
the core of the LOFAR telescope.

The EoR maps are simulated between the frequencies of $115~{\rm MHz}$
and $178.5~{\rm MHz}$ in steps of $0.5~{\rm MHz}$, corresponding to
redshifts between 11.5 and 6.5.
 
The mean of the EoR signal, foreground and noise maps at each
frequency are set to zero since LOFAR is an interferometric instrument
and measures only fluctuations around the mean. The typical
variations, $\sigma$, over the map at 150~{\rm MHz} are $\sim5~{\rm
mK}$ for the EoR signal, $\sim2~{\rm K}$ for the foregrounds and
$\sim52~{\rm mK}$ for noise. Hereafter, these values are considered
fiducial values for the EoR signal, foregrounds and noise.

\begin{figure}
\centering
\includegraphics[width=.45\textwidth]{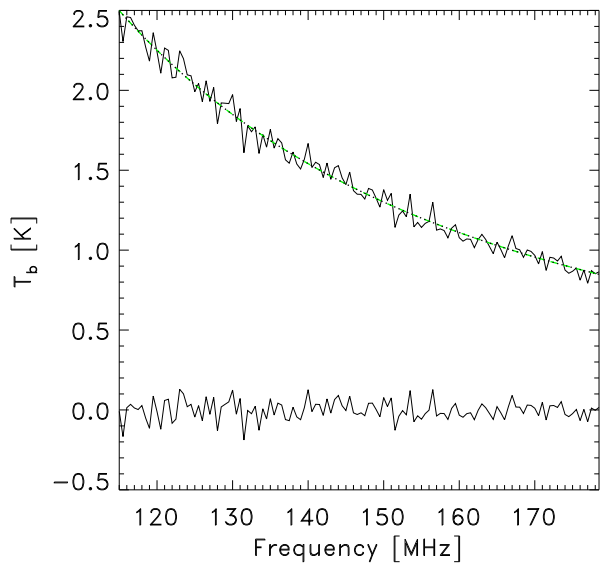}
\caption{\emph{One line of sight (one pixel along frequency) of the
`original' LOFAR-EoR data maps (upper solid black line), smooth
component of the foregrounds (dotted black line), fitted foregrounds
(dashed green line) and residuals (lower solid black line) after
taking out the foregrounds. Note that the residuals are the sum of the
EoR signal and the noise.}}
\label{pix}
\end{figure}

\begin{figure}
\centering
\includegraphics[width=.45\textwidth]{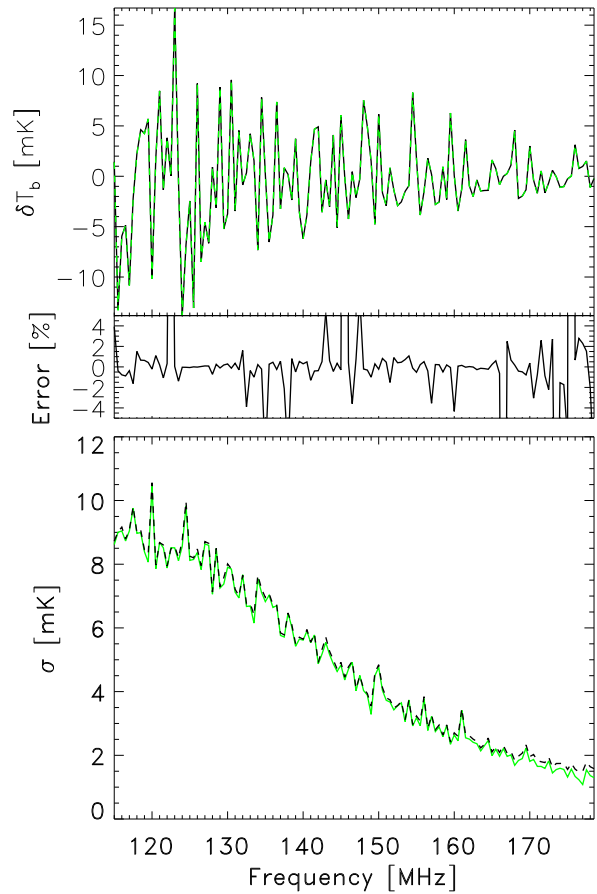}
\caption{\emph{Detection of the EoR signal from the simulated
foreground maps (`original' maps), without interferometric effects and
noise: for a single random line of sight (top panel) and as a standard
deviation over all lines of sight (bottom panel). The solid green line
represents the original simulated EoR signal, and the dashed black
line the extracted EoR signal.}}
\label{cs}
\end{figure}

The analysis on the LOFAR-EoR data maps can be done in two ways:
firstly along the frequency direction where the foregrounds are
assumed to be smooth in contrast to the EoR signal; and secondly in
the spatial domain where the EoR signal and some components of the
foregrounds are spatially correlated, but the noise is not. In this
paper we will demonstrate statistical detection of the signal by analysis
along the frequency direction, taking lines of sight (map pixels) one
by one.

Fig.~\ref{pix} shows one line of sight (one pixel along frequency) of
the `original' LOFAR-EoR data cube (upper black solid line) without
interferometric effects. The first step in the extraction of the EoR
signal is to take out the smooth foreground component (dotted black
line). It is important to note, however, that the smooth component of
the foregrounds is not a simple power law but a superposition of three
power laws (Galactic synchrotron and free-free emission and integrated
emission from unresolved extragalactic sources) including the fact
that one of the power law indices $\beta$ (Galactic synchrotron
emission) changes slightly with frequency.

The simplest method for foreground removal is a polynomial fitting in
logarithmic scale ($\log (T_{b})-\log (\nu)$). The dashed green line
on Fig.~\ref{pix} represents the foregrounds fitted with a 3$^{\rm
rd}$ order polynomial in the logarithmic scale.

Fig.~\ref{cs} shows a comparison between the detected and original EoR
signal for randomly chosen lines of sight in the case of the fiducial
foreground level and without noise for `original' maps. As one can
see, there is an almost exact agreement: this confirms that our
approach when applied to noiseless data does not introduce any
systematic biases.

After taking out the fitted foregrounds from the `original' data maps,
the residuals should contain only the noise and the EoR signal (lower
solid black line on Fig.~\ref{pix}).  However, the assumption here is
that we have fitted well enough such that the residuals between the
fitted and the `real' foregrounds are smaller than the EoR
signal. Otherwise the EoR signal could be fitted out if we are
over-fitting, or be dominated by the foreground fitting residuals if
we are under-fitting the foregrounds.

The extraction of the EoR signal from the residuals along one line of
sight is an impossible task, since the level of the noise is order of
magnitudes larger than EoR signal and its value is unknown for a
certain pixel. However, general statistical properties of the noise
(standard deviation as a function of frequency) might be determined
from the experiment and be used to statistically detect the EoR
signal. By statistical detection we mean determination of the standard
deviation of the EoR signal over the entire map as an excess variance
over the variance of the noise. The general statistical properties of
the noise might be determined in two ways. The first method is based
on the difference between measured fluxes of a discrete point source,
with well know properties, at two consecutive frequency channels. The
second one is based on the difference between the measured flux in
total and polarized intensity at the same frequency. However, the
accuracy of the both methods need to be tested for the LOFAR-EoR
experiment and we leave further discussion on this topic for a
forthcoming paper.

\begin{figure}
\centering \includegraphics[width=.45\textwidth]{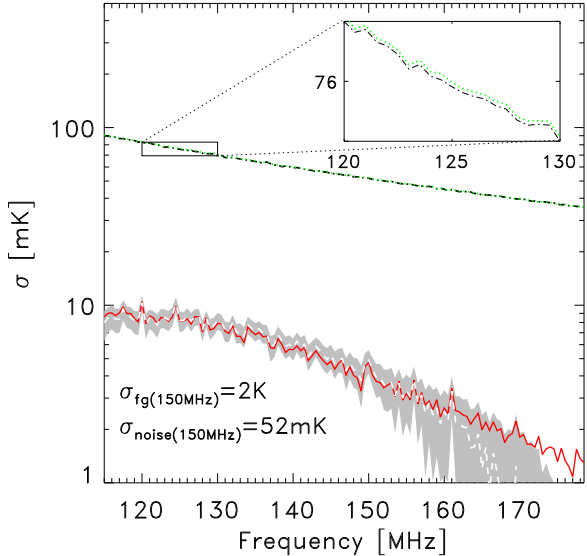}
\caption{\emph{Statistical detection of the EoR signal from the
`original' LOFAR-EoR data maps that include diffuse components of the
foregrounds and realistic noise ($\sigma_{\rm noise}(150~{\rm
MHz})=52~{\rm mK}$) but without interferometric effects. The
dashed-dotted black line represents the standard deviation ($\sigma$)
of the noise as a function of frequency, the dotted green line the
$\sigma$ of the residuals after taking out the smooth foreground
component, and the solid red line the $\sigma$ of the original EoR
signal. The grey shaded zone shows the $2\sigma$ detection, whereas
the dashed white line shows the mean of the detection. Note that the
y-axis is in logarithmic scale.}}
\label{nfg}
\end{figure}

Fig.~\ref{nfg} shows the standard deviation of residuals as a function
of frequency (dotted green line), after taking out the smooth
component of the foregrounds, by polynomial fitting in logarithmic
scale to each line of sight of the `original' maps. The most
satisfactory result we get with a 3$^{\rm rd}$ order polynomial. The
dashed-dotted black line represents the standard deviation of the
noise. By subtracting (in quadrature) the $\sigma_{\rm noise}$ from
$\sigma_{\rm residuals}$, we get the excess variance ($\sigma_{\rm
EoR}$) of the EoR signal. However, in order to determine the error on
the detection of the EoR signal, we conducted a Monte-Carlo simulation
of the extraction of the signal. We made 1000 independent noise cube
realisations and applied the signal extraction algorithm to each. The
results of the simulation are shown in Fig.~\ref{nfg}. The grey shaded
zone shows the 2$\sigma$ detection, whereas the dashed white line
shows the mean of the detection.  As one can see the mean of the
detected EoR signal is in good agreement with the original (solid red
line) up to 165~{\rm MHz}. The disagreement for higher frequencies is
due to over-fitting and low EoR signal level.  Remember that for most
of the sightlines our simulated Universe has already been ionized at
this frequency, corresponding to $z\approx 7.5$ (see
section~\ref{sec:signal}).

\begin{figure*}
\centering \includegraphics[width=.33\textwidth]{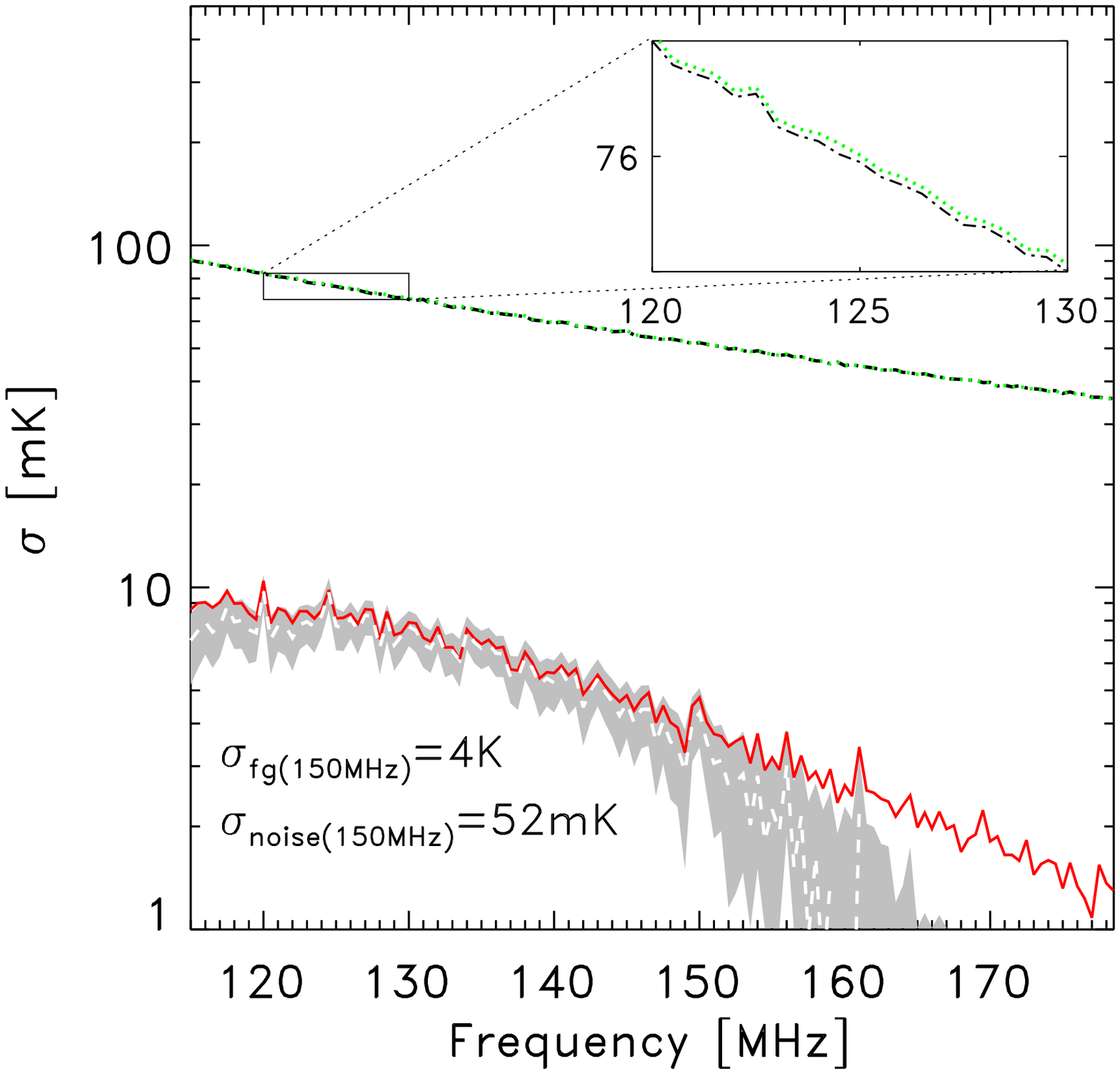}
\includegraphics[width=.33\textwidth]{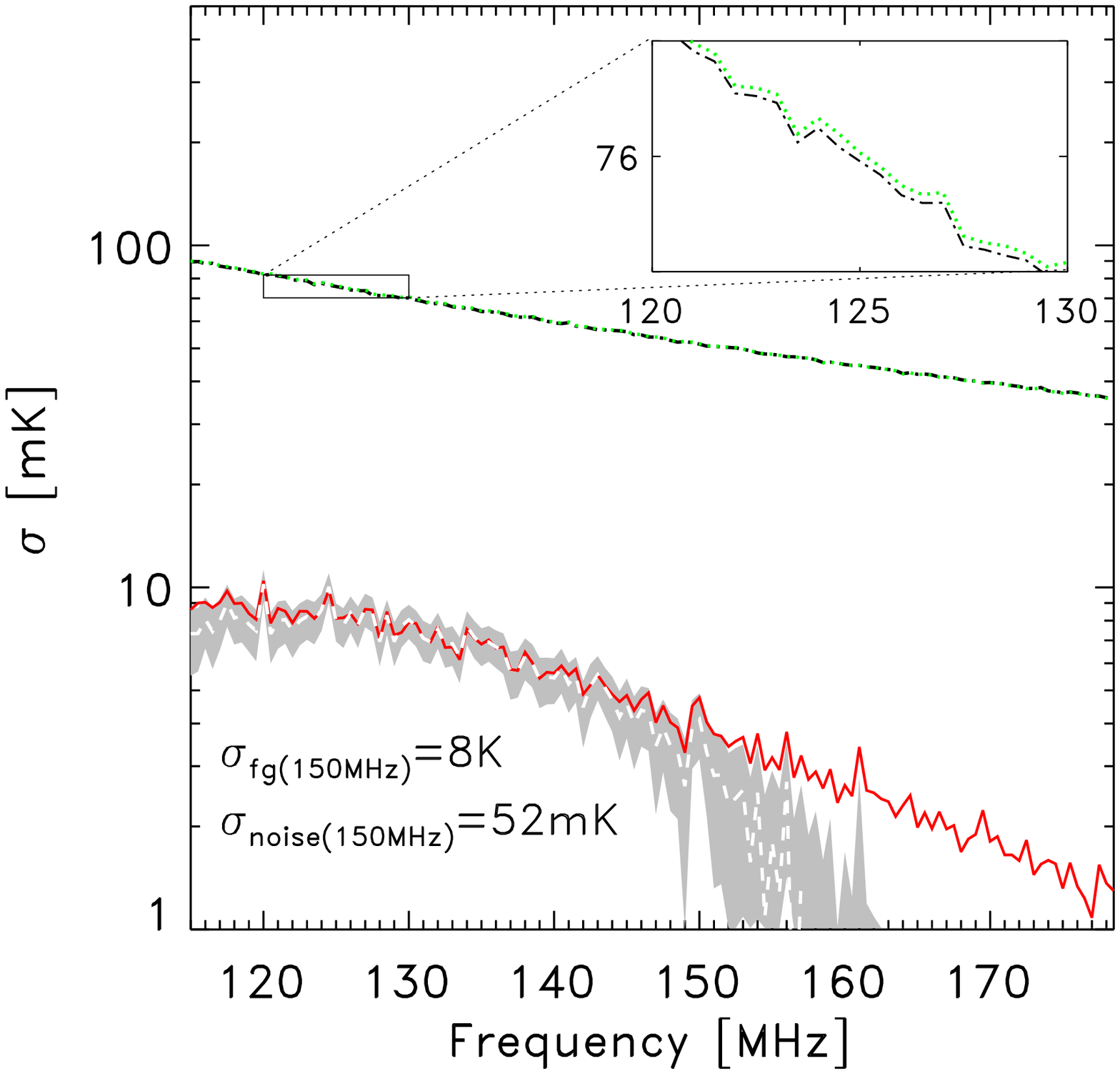}
\includegraphics[width=.33\textwidth]{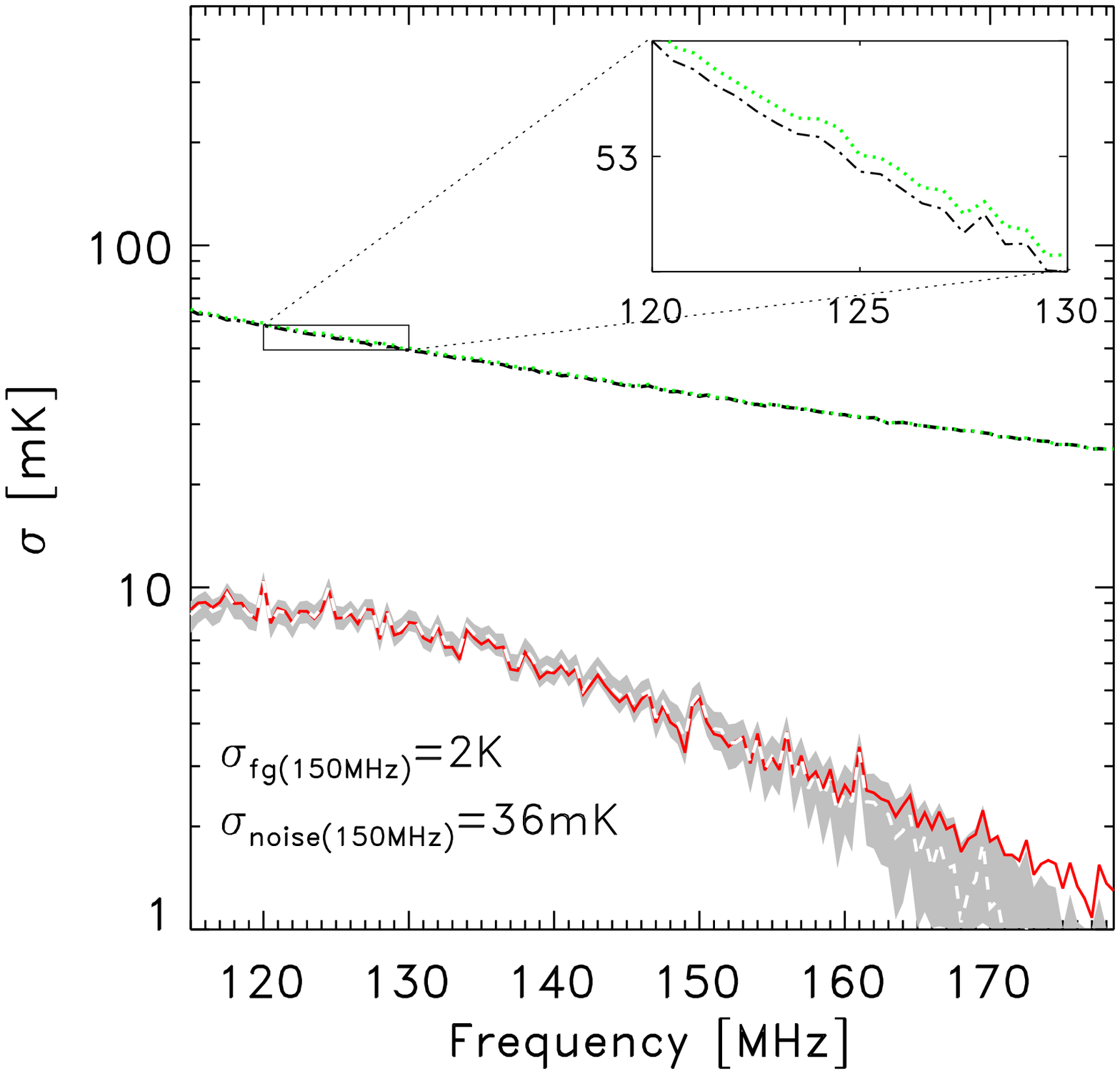}
\caption{\emph{Statistical detection of the EoR signal from the
`original' LOFAR-EoR data maps with foreground level two (left panel)
and four (middle panel) times bigger than the fiducial foreground
level, and with noise level smaller by $\sqrt{2}$ (right panel) than
the fiducial noise level, but without interferometric effects. Colours
and line coding are the same as in Fig.~\ref{nfg}. Note that the
y-axis is in logarithmic scale.}}
\label{nfg24}
\end{figure*}

In order to see the influence of the foreground and noise level on the
EoR extraction and detection scheme, we repeated the same analysis on
`original' maps of the four different models.  The first model has a
foreground level two times bigger than fiducial and the second four
times; the third has the fiducial foreground level but smaller noise
level by $\sqrt{2}$; and the last one has a normal foreground level
and no noise (see Table~\ref{models}). Note that by fiducial
foreground level and noise level we mean $\sigma_{fg}(150~{\rm
MHz})\sim 2~{\rm K}$ and $\sigma_{\rm noise}(150~{\rm MHz})\sim
52~{\rm mK}$.  The results are shown in Figs.~\ref{cs} \& \ref{nfg24}.

\begin{table}
  \centering
  \caption{Five different sets of values for standard deviation of
  foregrounds ($\sigma_{fg} [{\rm K}]$ at $150~{\rm MHz}$) and of
  noise ($\sigma_{noise} [{\rm mK}]$ at $150~{\rm MHz}$), used for
  testing the EoR extraction and detection scheme. Note that case (a)
  represents the fiducial case.}
  \begin{tabular}{@{}cccccc@{}}
    \hline & case (a) & case (b) & case (c) & case (d) & case (e)\\
    \hline $\sigma_{fg} [{\rm K}]$& \textbf{2} & 4 & 8 & 2 & 2\\
    $\sigma_{noise} [{\rm mK}]$& \textbf{52} & 52 & 52 & 36 & 0\\
    \hline
  \end{tabular}\label{models}
\end{table}

Comparing Figs.~\ref{nfg} \& \ref{nfg24}, we see the higher foreground
levels decrease the quality of the EoR detection. Lower quality in the
EoR detection is due to over-fitting at higher frequencies. However,
even for the four times bigger foreground level we are able to detect
the EoR signal up to $150~{\rm MHz}$.

Comparing Figs.~\ref{cs}, \ref{nfg} \& \ref{nfg24}, we can see that a
lower noise level increases the quality of the EoR detection, as
expected. Better precision in the EoR detection with lower noise level
also confirms that our foreground removal procedure works well.

\begin{figure}
\centering \includegraphics[width=.45\textwidth]{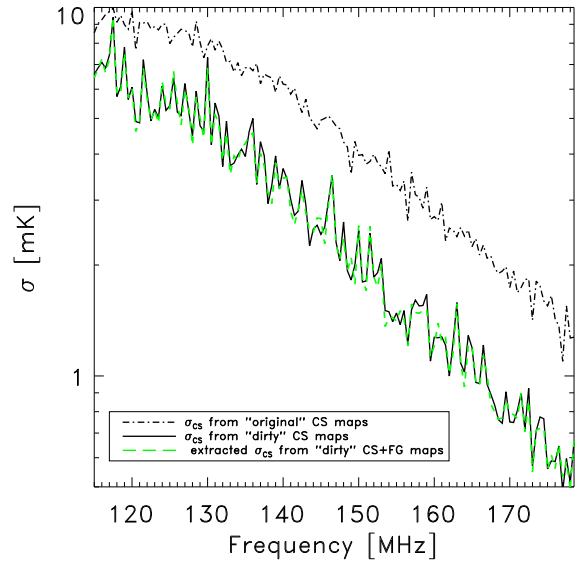}
\caption{\emph{Detection of the EoR signal from the simulated
simplified `dirty' foreground maps, without noise, as a standard
deviation $\sigma$ over all lines of sight (dashed green line). The
dashed-dotted black line represents the $\sigma$ over all lines of
sight of `original' EoR signal, while the solid black line the
$\sigma$ over all lines of sight of `dirty' EoR signal maps.}}
\label{stddm}
\end{figure}

Finally, in Fig.~\ref{stddm}, we show the statistical detection of the
EoR signal from the `dirty' foregrounds + EoR signal maps without any
noise. Note that the `dirty' maps are produced with a sampling
function (uv mask) that contains only uv points present at each
frequency, in order to overcome additional difficulties from the
mixing of angular scales in the frequency direction introduced by the
linear frequency variation of the uv coverage and incomplete sampling
in the frequency direction.

The smooth component of the foregrounds is removed by polynomial
fitting to each line of sight. The most satisfactory result we get
with a 6$^{\rm th}$ order polynomial. A different order of polynomial
from the case of the `original' maps is required due to the angular
scale mixing over each map introduced by convolution of the map with a
`dirty' beam.

\begin{figure}
\centering \includegraphics[width=.45\textwidth]{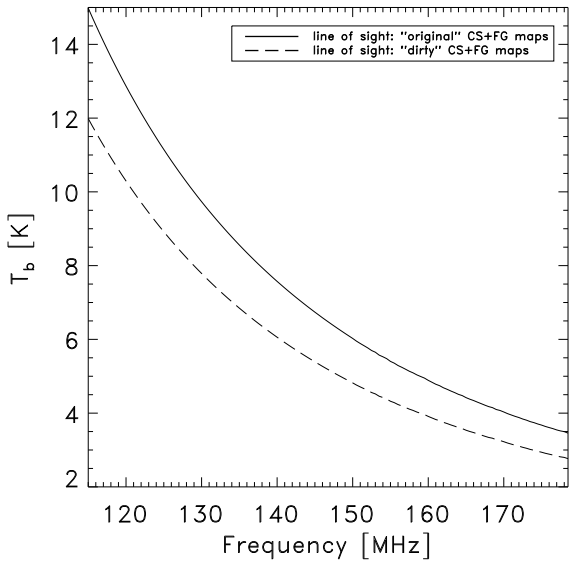}
\caption{\emph{One line of sight (one pixel along frequency) of the
`dirty' foreground (FG) + cosmological 21-cm signal (CS) maps (dashed
line) in comparison with the same pixel along the frequency of the
`original' FG+CS maps (solid line). The difference between these two
lines is due to incomplete uv coverage and its finite extent.}}
\label{dmpix}
\end{figure}

Fig.~\ref{dmpix} shows a comparison along the frequency direction of
the same pixel from the `original' (solid line) and `dirty' (dashed
line) foregrounds + signal maps. Note that the foregrounds are still
smooth along the frequency direction of the `dirty' maps, but the
shape of the function is slightly different. The difference is due to
incomplete uv coverage sampling and its finite extent, determined by
the shortest and longest baselines.

The dashed green line in Fig.~\ref{stddm} shows the standard
deviation, as a function of frequency, of the extracted EoR signal
from `dirty' foregrounds + EoR signal maps. The dashed-dotted black
line shows the standard deviation of the `original' EoR signal maps,
while the solid black line shows the standard deviation of the `dirty'
EoR signal maps. The agreement between the standard deviations of the
extracted and `dirty' EoR signals is satisfactory, while their
slightly lower levels than the `original' signal are due to the
smoothing effect of the instrumental response.

\section{Discussion and Outlook}
\label{sec:discussion and outlook}

This paper presents foreground simulations tailored for the LOFAR-EoR
experiment that is set to study the redshifted 21-cm hyperfine line of
neutral hydrogen from the Epoch of Reionization.  The foreground
simulations include Galactic diffuse synchrotron and free-free
emission, synchrotron emission from Galactic supernova remnants and
extragalactic emission from radio galaxies and clusters. For each of
the foreground components, we generate the $5^\circ\times 5^\circ$
field in the frequency range approximately between 115 and 180~{\rm
MHz} pertaining to the LOFAR-EoR.

Since the diffuse Galactic synchrotron emission is the dominant
component ($\sim 70$ per cent) we include all its observed
characteristics: spatial and frequency variations of brightness
temperature and its spectral index, and brightness temperature
variations along the line of sight. Discrete sources of Galactic
synchrotron emission are included as observed emission from supernovae
remnants.

Despite the minor contribution of the Galactic free-free emission
($\sim 1$ per cent), it is included in our simulations of the
foregrounds as an individual component. It has a different temperature
spectral index from Galactic synchrotron emission.

Integrated emission from extragalactic sources is decomposed into two
components: emission from radio galaxies and from radio
clusters. Simulations of radio galaxies are based on the source count
functions at low radio frequency by \citet{jackson05}, for three
different types of radio galaxies, namely FR{\small I}, FR{\small II}
and star forming galaxies. Correlations obtained by radio galaxy
surveys are used for their angular distribution. Simulations of radio
clusters are based on a cluster catalogue from the Virgo consortium and
observed mass--X-ray luminosity and X-ray--radio luminosity relations.

Under the assumption of perfect calibration, LOFAR-EoR data maps that
include the simulated cosmological 21cm signal ($\sigma_{\rm EoR}(150~{\rm
MHz})\sim 5~{\rm mK}$), diffuse components of the foregrounds
($\sigma_{FGs}(150~{\rm MHz})\sim 2~{\rm K}$) and realistic
noise ($\sigma_{\rm noise}(150~{\rm MHz})\sim 52~{\rm
mK}$) are produced. We refer to this set of parameters as our
fiducial model.  For noise we assume it has two
components, the sky noise and receiver noise. The former varies with
frequency as $\nu^{-2.55}$ whereas the latter is roughly frequency
independent.

The extraction of the EoR signal is performed along the frequency
direction, taking lines of sight (map pixels) one by one. The first
step in the EoR extraction is removal of the smooth foregrounds
component for each line of sight (see Fig.~\ref{pix}). In our analysis
we fit a 3$^{rd}$ order polynomial in the logarithmic scale. However,
one should be careful in choosing the order of the polynomial to
perform the fitting. If the order of the polynomial is too small, the
foregrounds will be under-fitted and the EoR signal could be dominated
and corrupted by the fitting residuals, while if the order of the
polynomial is too big, the EoR signal could be fitted out.

After foreground removal, the residuals are dominated by instrumental
noise. Since the noise is unknown for each line of sight and is an
order of magnitude larger than the EoR signal, it is an impossible task
to directly extract the EoR signal for each line of sight. However,
assuming that the statistical properties of the noise ($\sigma_{\rm
noise}$) will be known, we can use it to statistically detect the EoR
signal. The statistical detection of the EoR signal is the measure of
the excess variance over the entire map, $\sigma_{\rm EoR}^{2}$, that
should be obtained by subtracting the variance of the noise,
$\sigma_{\rm noise}^{2}$, from that of the residuals, $\sigma_{\rm
residuals}^{2}$.

Fig.~\ref{nfg} shows the results of a successful statistical detection
of the EoR signal in the fiducial model of the foregrounds and
noise. The detected standard deviation of the signal is in a good
agreement with original signal up to 165~{\rm MHz}. The disagreement
for higher frequencies is due to over-fitting caused by the very weak
cosmological signal at these frequencies.

In order to see the influence of the foreground and noise level on the
EoR extraction and detection, the same analysis was done for models
with two and four times bigger foreground levels than in the fiducial
model, for a model where the noise is smaller by $\sqrt{2}$, and for a
model without noise (see Tabel~\ref{models}). The results are shown in
Figs.~\ref{nfg24} \& \ref{cs}.

In the case of higher foreground levels, the EoR signal detection is
hampered by over-fitting. In the case of lower noise
levels, however, the proposed EoR detection algorithm performs extremely well.

For the diffuse components of simulated foregrounds, a `dirty' map
with realistic but idealised instrumental response of LOFAR is
produced (see Fig.~\ref{fig:dmmap}). However, the signal extraction
scheme we apply only to the `dirty' maps that have been produced with a
uniform uv coverage as a function of frequency and have no noise. This
is due to the additional difficulties introduced by mixing of angular
scales in the frequency direction. Those issues will be discussed in a
follow-up paper.

In addition to the simulations of the total brightness temperature,
the polarized Galactic synchrotron emission maps are produced. Here,
we follow a simple model that includes multiple Faraday screens along
the line of sight (see Figs.~\ref{fig:Iplot} \& \ref{fig:Qplot}). The
motivation behind these simulations is that improper polarization
calibration could severely contaminate the EoR signal, so future
robust extraction algorithms have to take this into account.

\begin{figure}
\centering
\hspace{-0.5cm}
\includegraphics[width=.5\textwidth]{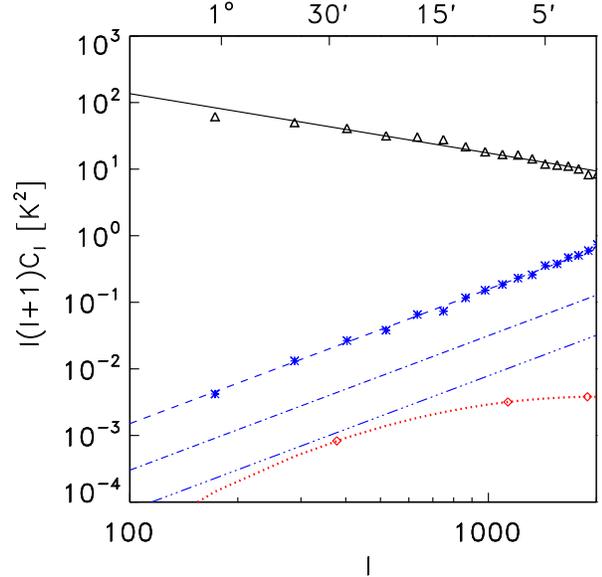}
\vspace{-1.cm}
\caption{\emph{Angular power spectra of the simulated EoR signal
(dotted red line), simulated dominant component of the foregrounds
(solid black line) and three levels of noise: the dashed blue line
represents noise for a single beam after one year of integration, the
dashed-dotted blue line for five beams after one year of integration
and the dashed-dotted-dotted blue line for five beams and four years
of integration. The lines are drawn as the best fit to the
coresponding points.}}
\label{fig:angps}
\end{figure}

Fig.~\ref{fig:angps} shows the angular power spectra of the simulated
EoR signal (dotted red line), simulated diffuse component of the
foregrounds (solid black line) and three levels of noise (blue lines)
at $150~{\rm MHz}$. The lines are drawn as the best fit to the
coresponding points.  The dashed blue line represents the level of the
noise ($\sigma_{\rm noise}(150~{\rm MHz})=52~{\rm mK}$) after one year
of the LOFAR-EoR experiment ($400~$h of total observing time) with a
single beam. For this case of instrumental noise and inclusion of
realistic diffuse foregrounds we have shown that we are able to
statistically detect the EoR signal despite the small signal to noise
ratio. However, the current observing plan of the LOFAR-EoR experiment
\citep{PP07} is to observe with five independent beams, which reduces
the $\sigma_{\rm noise}$ by a factor of $\sqrt{5}$ (dashed-dotted blue
line). After four years of observations ($4 \times 400~$h) with five
beams the $\sigma_{\rm noise}$ is reduced by a factor of $\sqrt{20}$
(dashed-dotted-dotted blue line), which means that the signal to noise
ratio is roughly 0.5.

Finally we would like to emphasize that this paper is just a first
step in testing and developing the analysis tools for the LOFAR-EoR
experiment. Future papers will use the foregrounds simulations
developed in this paper together with simulations of the EoR signal
\citep[][in prep.]{thomas08}, instrumental response \citep[][in
prep.]{panos08}, and other non-astrophysical effects
(e.g. ionosphere, RFIs, ...) in order to test all aspects of the
pipeline in the LOFAR-EoR experiment.

\section*{acknowledgement}
The authors thank to the anonymous referee for his illustrative and
constructive comments. As LOFAR members, the authors are partially
funded by the European Union, European Regional Development Fund, and
by `Samenwerkingsverband Noord-Nederland', EZ/KOMPAS.

\appendix

\bsp

\label{lastpage}

\end{document}